\documentclass{emulateapj}

\usepackage{xspace}
\usepackage{graphicx}
\usepackage{natbib}
\usepackage{apjfonts}
\usepackage{etoolbox}
\usepackage{subfigure, slashbox} 
\usepackage{fp,url}

\newcommand{\src}{4U~1957+11}
\newcommand{\chandra}{\textsl{Chandra}}
\newcommand{\xmm}{\textsl{XMM-Newton}}
\newcommand{\suzaku}{\textsl{Suzaku}}

\newcommand{\swift}{\textsl{Swift}}

\newcommand{\nh}{N\ensuremath{_{\rm H} }}	% N_H
\newcommand{\msun}{M\ensuremath{_\odot}}	% Solar Mass
\newcommand{\mdot}{\ensuremath{\dot m}}		% mass accretion rate
\newcommand{\about}{\ensuremath{\sim}}		% about
\newcommand{\csq}{\ensuremath{\chi^2}}		% chi-square
\newcommand{\astar}{\ensuremath{a^* }}		% a^*
\newcommand{\hd}{\ensuremath{h_d }}		% h_d
\newcommand{\lledd}{\ensuremath{L/L_{\rm Edd} }}% L/L_Edd
\newcommand{\mbh}{M\ensuremath{_{\rm BH} }}	% M_BH
\newcommand{\mdi}{\textbraceleft\mbh, 		% {M_BH, d, i}
\ensuremath{d}, \ensuremath{i}\textbraceright}
\newcommand{\webpage}{\url{http://dept.astro.lsa.umich.edu/~dmaitra/4u1957/}}

\newcommand{\inclval}{\ensuremath{77.6^{+1.5}_{-2.2} }}
\newcommand{\nhval}{\ensuremath{1.22^{+0.03}_{-0.06}\times10^{21}\;{\rm cm}^{-2}}}

\newcommand{\heatmap}[2]{
 \FPmul\xlf{1.000}{#1}
 \FPmul\xmd{0.796}{#1}
 \FPmul\xrt{1.130}{#1}

 \centering
 \includegraphics[height=\xlf\textwidth, trim = 0  0 0 90, clip, angle=-90]{M05.#2.d.ps}
 \includegraphics[height=\xmd\textwidth, trim = 0 55 0 90, clip, angle=-90]{M10.#2.d.ps}
 \includegraphics[height=\xrt\textwidth, trim = 0 55 0  0, clip, angle=-90]{M15.#2.d.ps}
}
\newcommand{\heatmapsimplelledd}[2]{
 \FPmul\xlf{1.000}{#1}
 \FPmul\xmd{0.852}{#1}
 \FPmul\xrt{1.098}{#1}

 \centering
 \includegraphics[height=\xlf\textwidth, width=2.5in, trim = 0  0 0 65, clip, angle=-90]{#2-M05.ps}
 \includegraphics[height=\xmd\textwidth, width=2.5in, trim = 0 40 0 65, clip, angle=-90]{#2-M10.ps}
 \includegraphics[height=\xrt\textwidth, width=2.5in, trim = 0 40 0  -10, clip, angle=-90]{#2-M15.ps}
}
\newcommand{\pxpy}[3]{
 \parbox{#3cm}{
  \centering
  \ifstrequal{#1}{#2}
  {\includegraphics[height=#3cm, trim= 0 0 0 0,angle=-90]
     {./mcmc_h1d_px00#1.ps}}
  {\includegraphics[height=#3cm, trim= 0 0 0 0,angle=-90]
     {./mcmc_h2d_px00#1py00#2.ps}}
 }
}
\slugcomment{Accepted for publication on August 18, 2014}
\shorttitle{Swift Monitoring of \src}
\shortauthors{Maitra et al.}

\begin{document}

\title{Results of the Swift Monitoring Campaign of the X-ray Binary \src: 
Constraints on Binary Parameters}

\author{% /*{{{*/
 Dipankar Maitra\altaffilmark{1,2}, 
 Jon M. Miller\altaffilmark{2},
 Mark T. Reynolds\altaffilmark{2},
 Rubens Reis\altaffilmark{2}, and
 Mike Nowak\altaffilmark{3}
} 
\affil{Department of Physics \& Astronomy, Wheaton College,
  Norton, MA 02766, USA}
\affil{Department of Astronomy, University of Michigan,
  Ann Arbor, MI 48109, USA}
\affil{Massachusetts Institute of
  Technology, Kavli Institute for Astrophysics, Cambridge, MA 02139,
  USA}
\email{maitra\_dipankar@wheatoncollege.edu}
% /*}}}*/

\begin{abstract}   % Abstract and keywords /*{{{*/
We present new results of uniform spectral analysis of \swift/XRT
observations of the X-ray binary system \src. This includes 26
observations of the source made between MJD 54282--55890 (2007 July 01 --
2011 November 25).  All 26 spectra are predominantly thermal, and can be
modeled well with emission from an accretion disk around a black hole.
We analyze all 26 spectra jointly using traditional \csq\ fitting as
well as Markov Chain Monte Carlo simulations.  The results from both
methods agree, and constrains on model parameters like inclination,
column density, and black hole spin.  These results indicate that the
X-ray emitting inner accretion disk is inclined to our line-of-sight
by \inclval\ degrees.  Additionally, the other constraints we obtain on
parameters like the column density and black hole spin are consistent with
previous X-ray observations.  Distances less than 5 kpc are unlikely and
not only ruled out based on our analysis but also from other independent
observations.  Based on model-derived bolometric luminosities, we require
the source distance to be $>$10 kpc if the black hole's mass is $>$10
\msun. If the hole's mass is $<$10 \msun, then the distance could be in
the range of 5--10 kpc.
\end{abstract}

\keywords{accretion, accretion disks --- binaries: general --- X-rays: 
binaries --- X-rays: individual: individual (\src)}
% End abstract and keywords /*}}}*/

\section{Introduction} \label{s:intro}  % /*{{{*/
\src\ is a bright, persistent X-ray source with soft X-ray (2--12
keV) flux levels between 20--70 milli-Crab since its discovery in
1973 \citep{giacconi:74a}.  Yet surprisingly little is known about
this binary system. While the lack of X-ray eclipses in this system
suggests that the orbital inclination is likely less than 85\degr,
modeling the optical modulation gives only a weak constraint of
$\sim20\degr<i<70\degr$ \citep{Mason+2012}.  Neither the distance to
the system nor the accretor's mass is well known.  However, examining
the equivalent width of the Ne\,{\sc ix} 13.45\AA\ line created
in the ISM, \citet{Nowak+2008} and \citet{Yao+2008} have suggested
a minimum distance of 5 kpc.  In the absence of any dynamical mass
measurement, analysis of X-ray/optical data from the source at various
points of time have suggested that it could either be a neutron star
\citep{yaqoob:93a,Singh+1994,ricci:95a,robinson:12a} or a black hole
\citep{Wijnands+2002,Nowak+2008,Nowak+2012}.  The morphology of the
optical light curve of \src\ varies with time.  However observations
densely sampled in time reveals a modulation with 9.33 hour period,
which is usually thought to be the orbital period of this system
\citep{Thorstensen1987, Bayless+2011, Mason+2012}.

The column density (\nh) in the direction of \src\ is quite small
(1--2$\times$$10^{21}$ atoms cm$^{-2}$), providing a clear view of the
disk.  Furthermore, the high-resolution grating data obtained by \chandra\
and \xmm, and analyzed by \citet{Nowak+2008}, show only absorption lines
due to the ISM and no lines intrinsic to the source.

\citet{Nowak+2008} have analyzed the entire set  of {\it Rossi
X-ray Timing Explorer} (RXTE) observations, as well as \chandra\ and
\xmm\ observations of this source. More recently \citet{Nowak+2012}
have also presented their analysis of \suzaku\ data of this source.
The predominantly soft spectrum and very low fractional variability
\citep{NowakWilms1999,Wijnands+2002,Nowak+2008} are characteristic of
\src\ being in a canonical soft state.  Recent radio non-detection of
\src\ with an upper limit of of 11.4 $\mu$Jy/beam using the {\it Jansky
Very Large Array} (JVLA) at 5--7 GHz by \citet{Russell+2011} is also
consistent with the prevalent wisdom, a.k.a. the jet--disk paradigm, where
jet production is strongly quenched in sources that are in a soft state.

Recent works critically examining the X-ray spectral and timing
properties using \chandra, \xmm, and RXTE \citep{Nowak+2008,Nowak+2012}
have suggested that the system harbors a black hole, and that it may be
the fastest spinning hole known so far.  In this work we {\it assume}
that the accretor is a BH and test the validity of this assumption under
a wide range of plausible parameter space.

We present the details of the \swift\ observations and data analysis in
\S\ref{s:data}. We then discuss spectral modeling in \S\ref{s:analysis},
starting with simple, phenomenological accretion disk plus power law
models in \S\ref{ss:diskbb_po} and then moving towards more physically
motivated disk models in \S\ref{ss:kerrbb}.  Joint analysis of all the
observations using traditional \csq\ fitting technique is presented
in \S\ref{ss:chisq} and that using Markov Chain Monte Carlo (MCMC)
simulations is presented in \S\ref{ss:mcmc}.  Finally, our conclusions
are summarized in \S\ref{s:conclusion}.

% End intro /*}}}*/

\section{\swift\ monitoring campaign of \src} \label{s:data} % /*{{{*/
As part of the \swift\ observatory's \citep{Gehrels+2004} {\it Guest
Observing} program number 7100116, \src\ was observed 21 times between
MJD 55700--55890 (2011 May 19 --2011 November 25).  Prior to this
monitoring campaign \src\ was also observed 5 times with \swift.
With an average flux of $\sim1.5\times10^{-9}$ erg/s/cm$^2$,
the source is quite bright in the X-ray telescope's bandpass (XRT,
\citealt{Burrows+2005}). Thereofre these observations were carried out
in {\it windowed timing} (WT) mode to avoid pileup.  The observation
logs are presented in Table~\ref{tab:obslog}.

The data extraction and reduction were performed using the
\textsc{Heasoft} software (v6.12) developed and maintained by NASA's
High Energy Astrophysics Science Archive Research Center (HEASARC).
We followed the extraction steps outlined in \citet{ReynoldsMiller2013}.
The raw data were reprocessed using the \texttt{xrtpipeline} command to
ensure that the latest instrument calibrations and responses were used.
Since data were collected in WT mode, events were extracted from
a rectangular region containing the source. Neighboring source-free
regions were used to extract background spectra. The \texttt{xrtexpomap}
task was used to generate exposure maps which were then applied to the
extracted data.  While we used the response matrices (RMF) supplied with
the latest calibration database, custom ancillary response function (ARF)
files for every observation were created using \texttt{xrtmkarf} task.
As per the Swift XRT CALDB Release Note\footnote{Released 2011 July 25;
URL:http://www.swift.ac.uk/analysis/xrt/files/SWIFT-XRT-CALDB-09\_v16.pdf}
a systematic error of 3\% was added to the spectra using the
\texttt{set\_sys\_err\_frac} command in \textsc{ISIS}.

% End s:data /*}}}*/

\section{Modeling the XRT spectra} \label{s:analysis} % /*{{{*/

\subsection{Phenomenological accretion disk plus power law models} % /*{{{*/
\begin{figure} % dbb kTin and normalization /*{{{*/
\centering
 \includegraphics[height=0.49\textwidth, angle=-90]{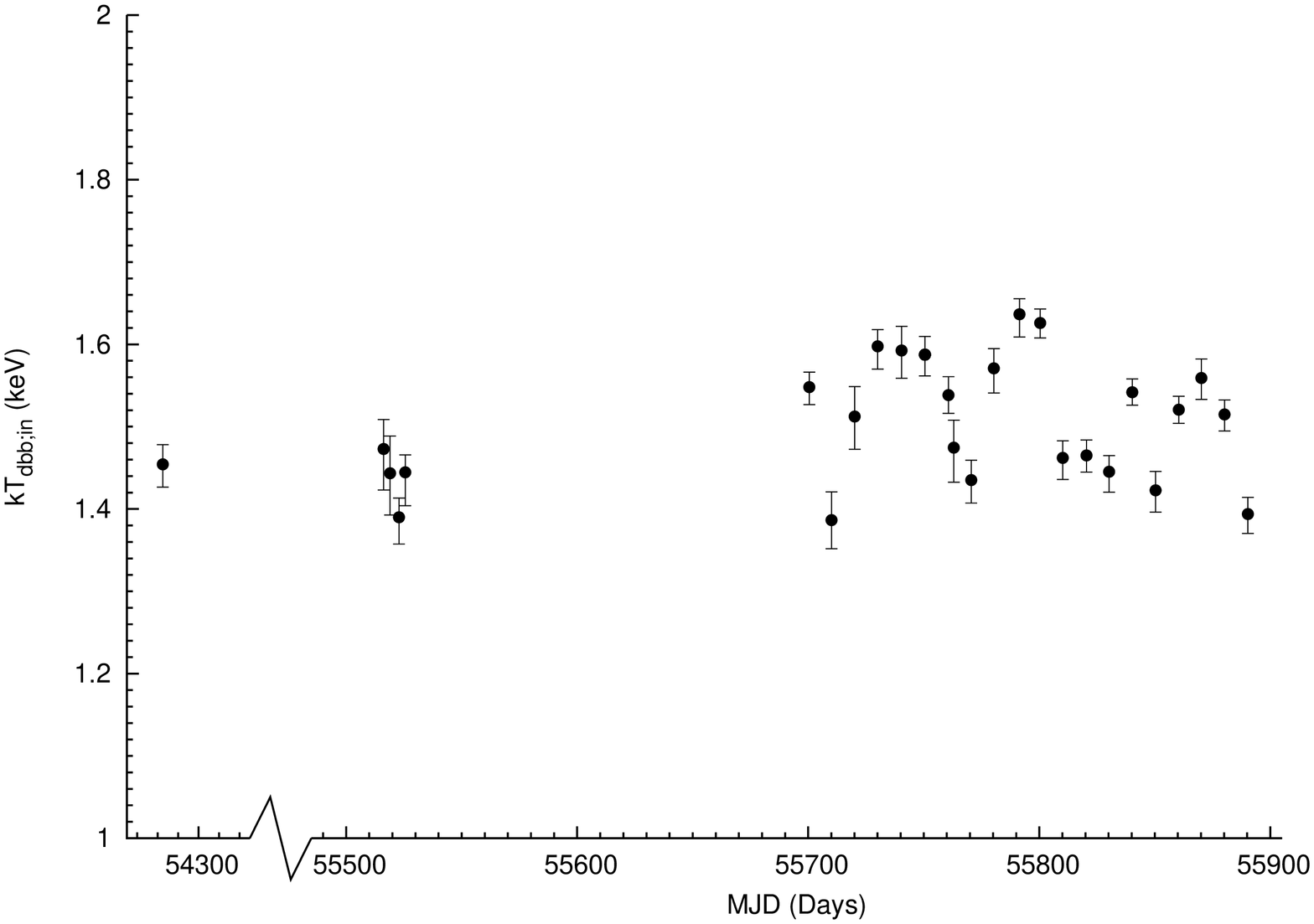} 
 \includegraphics[height=0.49\textwidth, angle=-90]{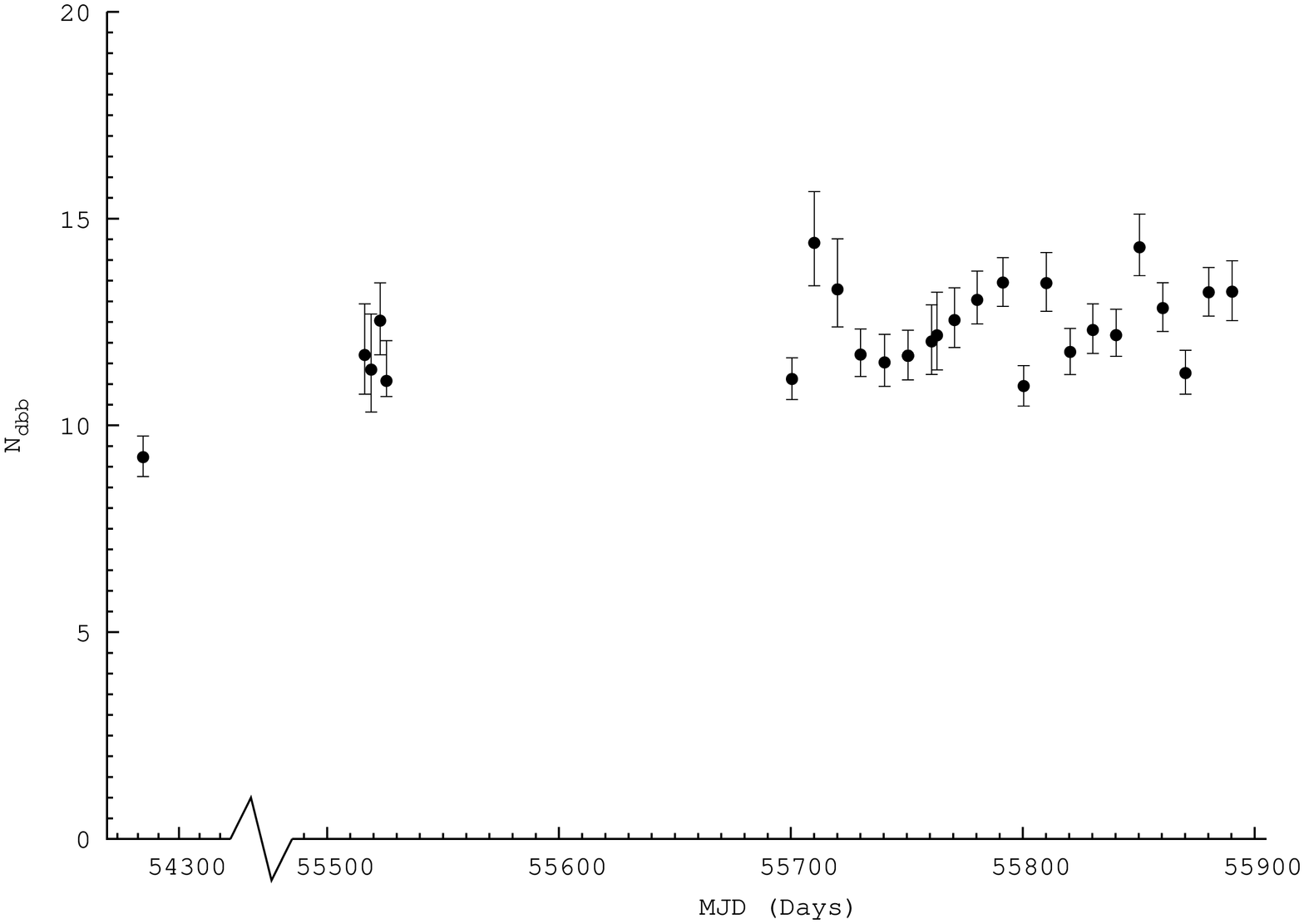} 
 \includegraphics[height=0.49\textwidth, angle=-90]{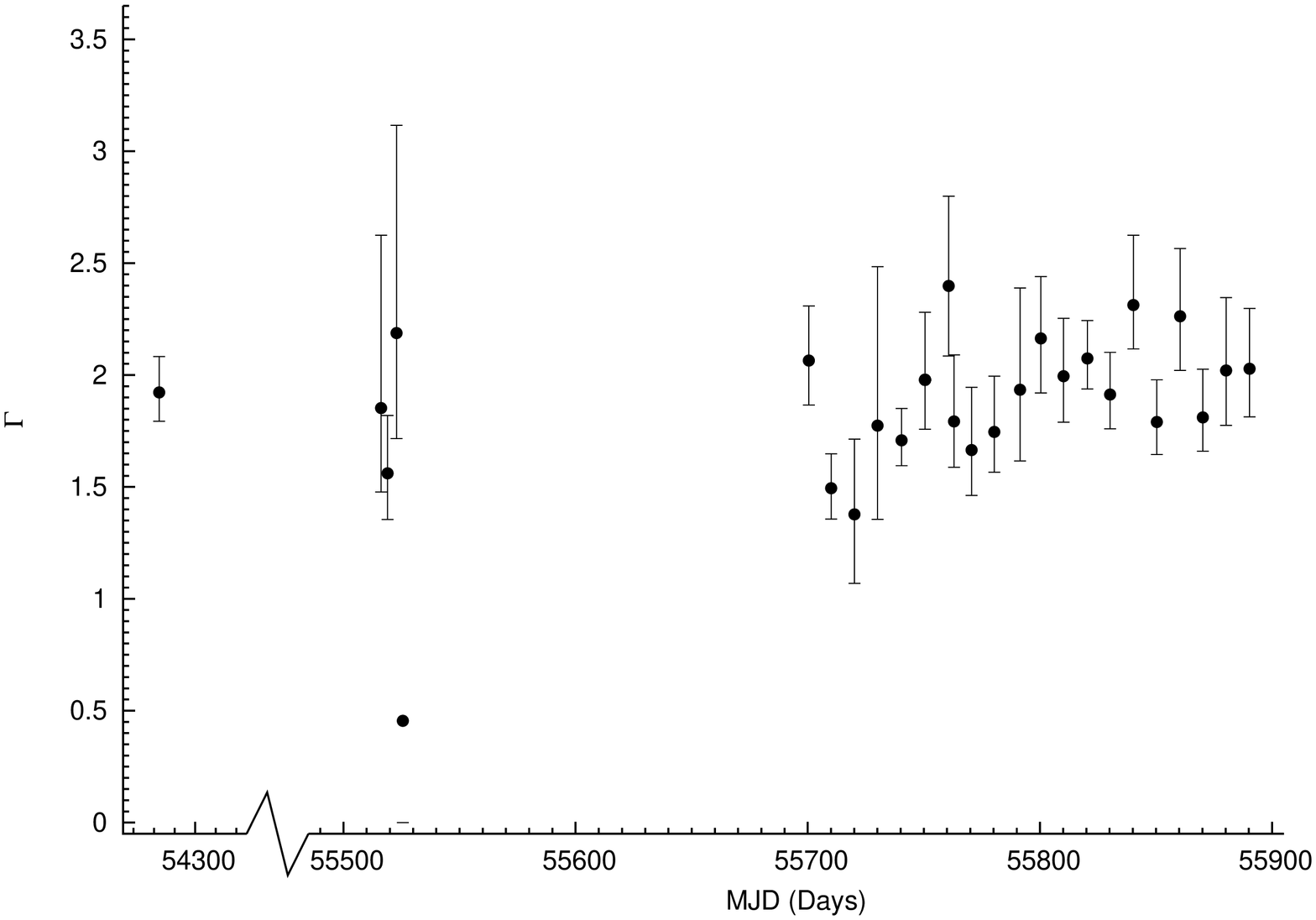} 
 \includegraphics[height=0.49\textwidth, angle=-90]{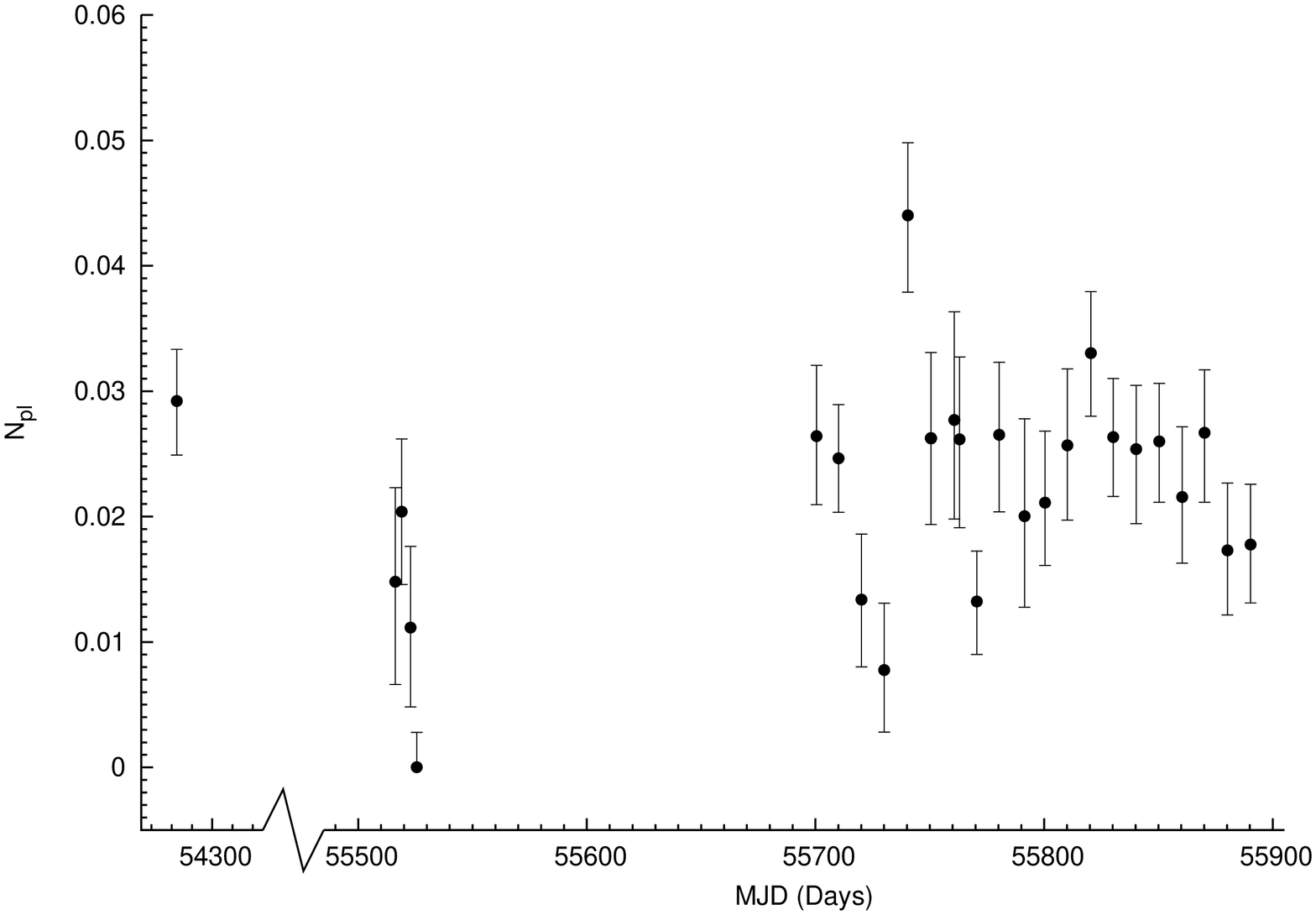} 
   \caption{{\em Top panels:}
   Best fit temperatures at the inner edge of the disk 
   ($kT_{\rm dbb;in}$), and disk normalizations ($N_{\rm dbb}$) for the 
   \texttt{diskbb+powerlaw} models.
   \newline {\em Bottom panels:} 
   Best fit photon indices ($\Gamma$) and normalizations of the 
   power law ($N_{\rm pl}$) for the same \texttt{diskbb+powerlaw} models.
   }
 \label{f:diskbb_po_diskpars} 
\end{figure}

\begin{figure} %%% Fluxes /*{{{*/
\centering
\includegraphics[height=0.49\textwidth, angle=-90]{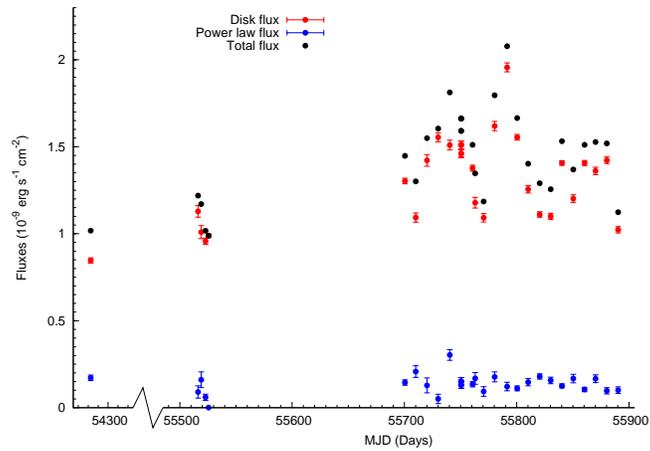} 
\caption{Unabsorbed fluxes from the best-fit phenomenological 
\texttt{diskbb+powerlaw} models.
}
\label{f:diskbb_po_fluxes} 
\end{figure}

\label{ss:diskbb_po}
In the simplest case we model the spectra with the standardly used
multi-temperature thermal accretion disk \citep[{\texttt{diskbb}}
in \textsc{XSPEC};][]{mitsuda:84a} plus a power law component
(\texttt{powerlaw} in \textsc{XSPEC}), modified by photoelectric
absorption (\texttt{phabs} in \textsc{XSPEC}) due to atoms
present in the intervening interstellar medium (ISM). We used the
\citet{andersgrevesse1989} abundance table and \citet{bcmc1992}
photoelectric absorption cross-sections (with a new He cross-section
based on \citet{Yan+1998}) to compute the photoelectric absorption
spectra.  The results of this spectral decomposition are shown
in Figs.~\ref{f:diskbb_po_diskpars}--\ref{f:diskbb_po_fluxes},
and summarized in Table~\ref{tab:obslog}. The fit parameters we
obtain are quite similar to the numbers obtained previously by
\citet{NowakWilms1999},\citet{Wijnands+2002}, and \citet{Nowak+2008}
during their analysis of the \textsl{RXTE} data of this source.
The spectra are predominantly thermal.  In fact, for the observation
on MJD 55525 which was the last of a batch of pointings between MJD
55516--55525, the disk plus power law decomposition fails to find any
nonthermal contribution.
% /*}}}*/

\subsection{Thin, thermal, relativistic accretion disk models around a Kerr black hole} % /*{{{*/
\label{ss:kerrbb}
The remarkable combination of low column density and absence of narrow
spectral features led \citet{Nowak+2008} to conclude ``\src\ may be
the cleanest disk spectrum with which to study modern disk atmosphere
models''.  We therefore used a second set of models where the observed
emission in the {\it Swift} bandpass (and given the limited spectral
resolution of the CCD chip) was entirely attributed to a thermal accretion
disk. As in the case of the phenomenological model described above,
we assumed that the intrinsic spectrum was modified by photoelectric
absorption (\texttt{phabs}) along the way.  For the accretion disk we
used the \texttt{kerrbb} model by \citet{Li+2005} which models a thin,
steady state, general relativistic accretion disk around a Kerr black
hole. While we encourage the reader to refer to the original work for
the details of the model, we give a brief summary of the relevant model
parameters here to provide a context.
%
% Start kerbb model description /*{{{*/
%

The model normalization was set to unity because the disk's inclination
($i$), the black hole's mass (\mbh), and the source distance ($d$)
were frozen during the analysis. The flags to switch effects of
self-irradiation and limb-darkening were turned on for all the fits. The
ratio of the disk power produced by torque at the disk's inner boundary
to the disk power arising from accretion ($\eta$) was set to 0 which
corresponds to a standard Keplerian disk with zero torque at the inner
boundary.  The above parameters were fixed for all fits.

The black hole's dimensionless spin parameter ($\astar$) was determined
from the fits. Similarly the mass accretion rate of the disk (\mdot),
and the spectral hardening factor $\hd$=$T_{col}/T_{eff}$ were also
determined via fitting. Here $T_{col}$ is the color temperature inferred
from the spectra and $T_{eff}$ is the effective temperature. As discussed
in greater detail in \citet{ShimuraTakahara:1995}, the spectral hardening
factor essentially parametrizes the uncertainties in our understanding
of the disk atmosphere.  Previous works on other sources (e.g., see,
\citet{Li+2005}, \citet{Shafee+2006}, \citet{McClintock+2006}) prefer
$\hd\sim1.7$ \citep[though also see][for fits to data where a higher \hd\
is preferred]{ReynoldsMiller2013, Salvesen+2013}.
% End --- kerbb model description /*}}}*/
% End -- kerrbb subsection /*}}}*/

\subsection{Traditional \csq\ fitting} % /*{{{*/
\label{ss:chisq}

For traditional \csq\ fitting we chose multiple \mdi\ triplets spanning
a wide range of masses (5, 10, and 15 \msun), distances (5, 10, 15,
and 20 kpc), and inclinations (55\arcdeg, 65\arcdeg, 75\arcdeg, and
85\arcdeg). Thus a total of 48 \mdi\ triplets were explored.
Inclinations lower than 55\degr\ result in fits that are progressively
worse fits, and therefore not considered.  The absence of X-ray eclipses
put an upper limit of 85\degr\ on the inclination. Similarly, examining
the equivalent width of the Ne\,{\sc ix} 13.45\AA\ line created in the
ISM, \citet{Nowak+2008} and \citet{Yao+2008} have estimated a minimum
distance of 5 kpc to this source.
While fitting any of these triplets, the values of \mbh, $d$, and $i$
were also kept constant. Thus the only free disk parameters in our
modeling are \astar, \hd, and \mdot.

However, the values of these free parameters are not completely
unconstrained: it is extremely unlikely that \astar\ changed appreciably
over the \about year timescale of the observations.  Therefore for our
joint fits (see details below) the value of \astar\ was tied to be the
same across all observations. In other words, the best-fit value of
\astar\ was determined from the data, but this value was required to be
the same for all observations.  As in the case of \astar, it is again
extremely unlikely that the column density of the intervening material
changed over a timescale of years.  Therefore the column density of
hydrogen (\nh\ in \texttt{phabs}) was also tied to be the same in
all observations.  Thus the only disk parameters that were free from
observation to observation were \mdot\ and \hd.

For every \mdi\ triplet we performed a joint fit to all 26 \swift\
observations. This implies that for every fit there were 54 free
parameters whose best-fit values were determined by fitting (one value
of \mdot\ and one value of \hd\ for every observation $\Rightarrow$
52 parameters, plus one value each of \nh\ and \astar). We rebinned the
spectra such that every bin had a signal-to-noise (S/N) ratio of at least
4.5, and used good data in the 0.5--10 keV energy range where calibration
of \swift/XRT's CCD is best known. After rebinning and energy filtering,
a total of 14,876 spectral bins (from all 26 observations) were used
for joint spectral fitting. Therefore the number of degrees of freedom
($\nu$) while jointly fitting all the observations for a given \mdi\
triplet is $\nu$=14,822.

Traditional $\chi^2$ minimization was carried out using the \textsc{ISIS}
software package \citep[v1.6.2-18][]{HouckDenicola2000}. \textsc{ISIS}
not only loads the entire library of models included in the
\textsc{XSPEC} \citep{Arnaud1996} package, but also allows
parallelized fitting and distributed computation of single-parameter
confidence limits in a cluster environment \citep[see, e.g.,][for
details]{Noble+2006,Maitra+2009}. Further speedup is gained by using
model caching methods in \textsc{ISIS} so that computationally expensive
models are not recomputed unless needed.

We used the \texttt{galaxy} cluster at the University of Michigan
to carry out fitting and distributed computation of single-parameter
confidence limits.  Using the parallelization scheme outlined above, we
determined the best-fitting parameter values as well as their confidence
intervals. Our confidence intervals correspond to  $\Delta\chi^2$= 2.71
for a given parameter (for normal distribution this would imply a single
parameter confidence limit of 90\%).
% End --- kerrbb fitting subsection /*}}}*/

\subsubsection{Results of \csq\ fitting} %/*{{{*/
\begin{figure} % Sample fits results from (10msun, 10kpc fits) /*{{{*/
\centering
 \begin{tabular}{|c|c|}
  \hline
  \includegraphics[height=0.45\textwidth, angle=-90]{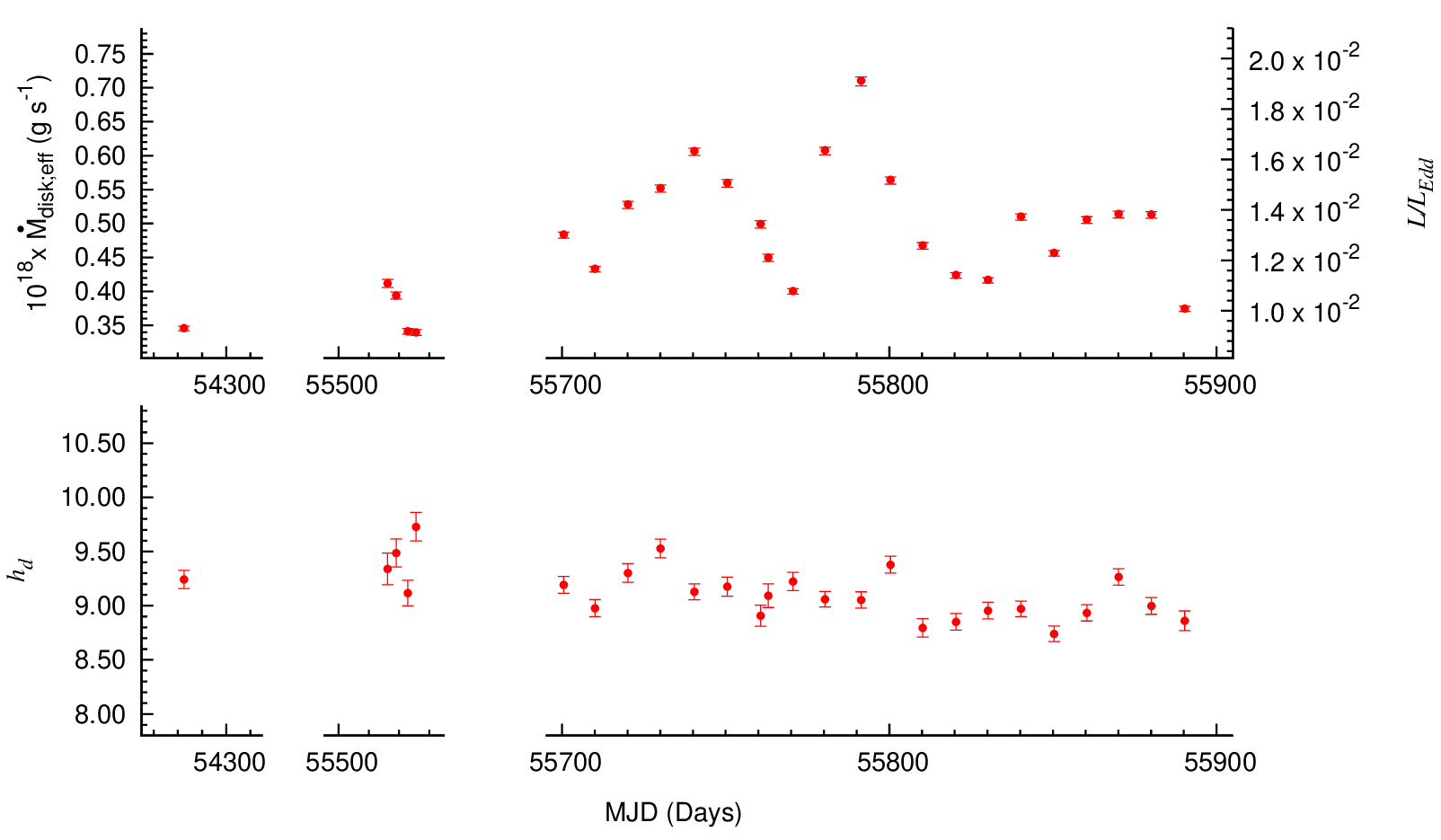} &
  \includegraphics[height=0.45\textwidth, angle=-90]{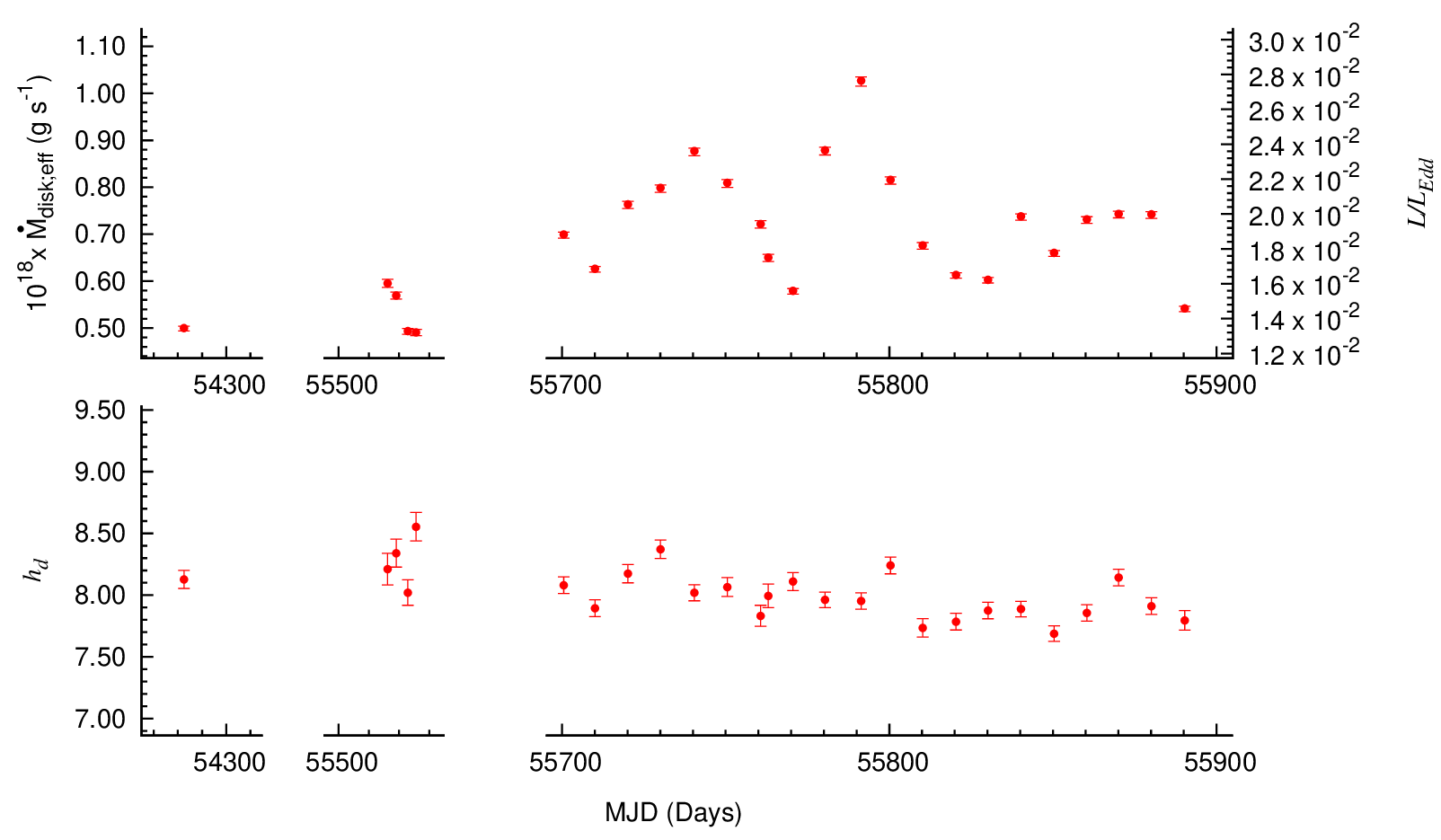} \\
  \hline
  \includegraphics[height=0.49\textwidth, angle=-90]{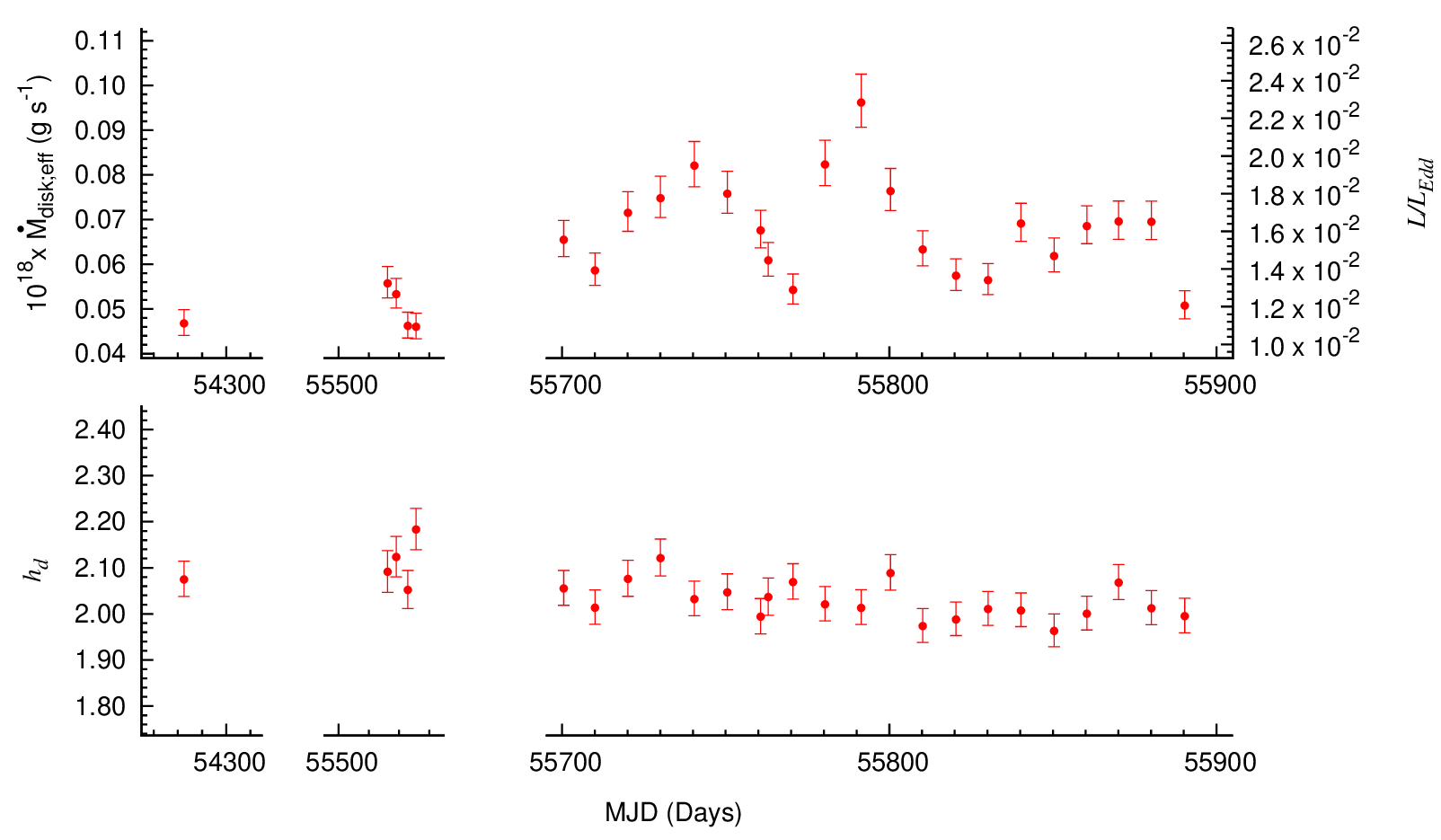} &
  \includegraphics[height=0.49\textwidth, angle=-90]{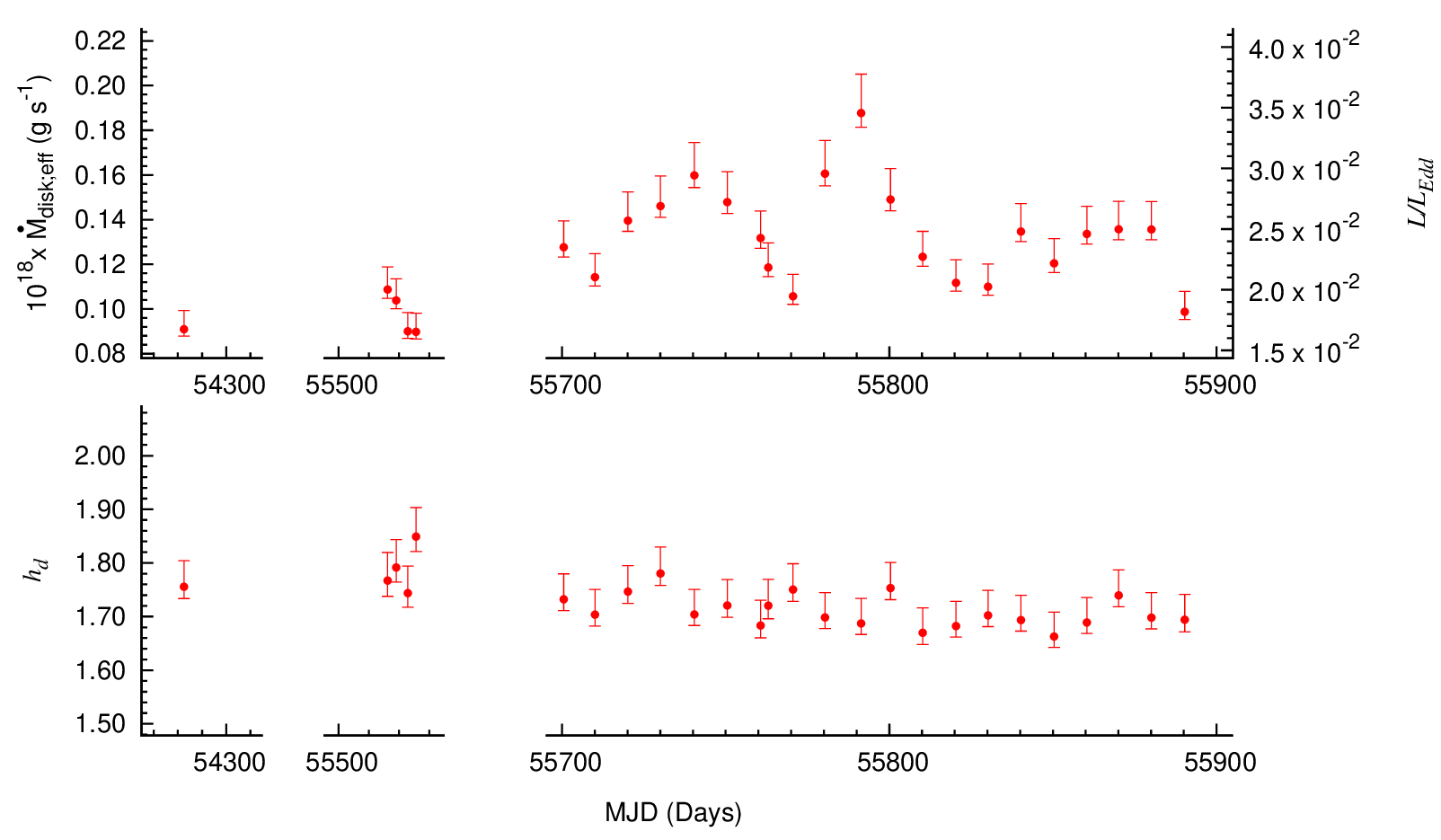} \\
  \hline
 \end{tabular}
 \caption{Best-fit values assuming $M$=10\msun, $D$=10 kpc, and
 $i$=55\arcdeg (top-left), 65\arcdeg (top-right), 75\arcdeg (bottom-left),
 and 85\arcdeg (bottom-right).  The top subpanel of each triplet shows the
 variation of \mdot\ and Eddington fraction (\lledd) with time,
 and the bottom subpanel shows the variation of spectral hardening factor
 (\hd). The best-fit \hd\ values for $i$=55\arcdeg\ and 65\arcdeg\ are
 clearly unphysical. The full set of figures showing
 the best-fit values as well as fitted and residual spectra for every
 observation for all the \mdi\ triplets explored in this paper can be
 seen online.
 }
 \label{f:M10_D10} 
\end{figure}

Given the plethora of model fits (48 \mdi~triplets, and 26 spectra
for each triplet $\Rightarrow$ 1248 spectral fits), we include only a
sample of fit results in this paper (see, e.g., Fig.~\ref{f:M10_D10}).
The complete set of results including
(a) best-fit model parameters for every \mdi~triplet, 
(b) best-fit spectral models for every observation showing the data, 
model, and residuals, and 
(c) time variation of \mdot\ and \hd\ for every \mdi\ triplet,
are available online\footnote{At \webpage}.

\begin{figure} % Heat maps based on chi-squared values /*{{{*/
 \heatmap{0.33}{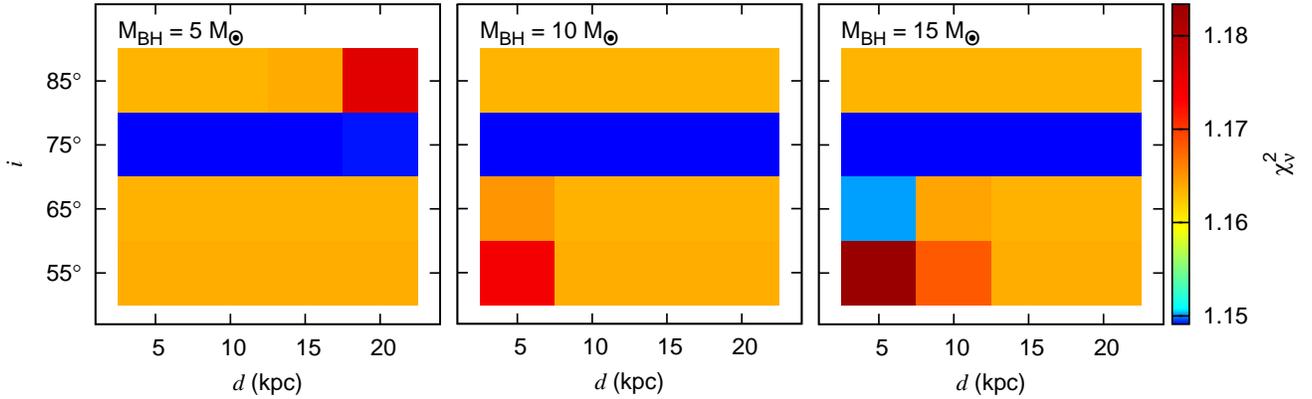}
   \caption{Heat maps based on best-fit reduced-$\chi^2$ values. 
   Each \mdi~triplet is represented by a square whose color 
   is indicative of the best-fit reduced-$\chi^2$ obtained 
   by simultaneously fitting all 26 observations. 
   The left, middle and right panels are
   for $M=5,10,15$\msun\ respectively. Note that an inclination of
   $\sim$75\arcdeg\ is preferred irrespective of the black hole's
   assumed mass.
   }
 \label{f:heat-chisq} 
\end{figure}

For easy exploration of the results we have created heat maps that show
the changes in fit statistics as well as best-fit parameters for the
different \mdi\ triplets. Fig.~\ref{f:heat-chisq} shows the heat maps
based on best-fit reduced-$\chi^2$ values\footnote{Since all \mdi\
triplets have 14822 degrees of freedom, any of the reduced $\chi^2$
values can be multiplied by this factor to obtain the actual $\chi^2$
value}. Each \mdi\ triplet is represented by a square whose color is
indicative of the best-fit reduced-$\chi^2$ obtained by simultaneously
fitting all 26 observations. Note that an inclination of \about75\arcdeg\
is statistically preferred irrespective of the black hole's mass.

Since we assume a standard Keplerian disk with zero torque at the
inner boundary, the total disk luminosity for the models is given by
$L=\epsilon\mdot c^2$.  Here $\epsilon$ is the radiation efficiency of the
disk around the accretor and depends on the spin (see, e.g. Figure 4 of
\citealt{Li+2005}). This allows us to calculate the fractional Eddington
luminosity from the model fits.  Since the \texttt{kerrbb} model takes
into account relativistic effects like Doppler beaming, deflection of
light under strong gravity (and the resulting ``returning radiation''),
and gravitational redshift, as well as incorporating additional physics
such as limb-darkening and self-irradiation of the disk, this estimation
of \lledd\ is significantly better than simply estimating it from the
observed flux.  Thus, e.g. in Fig.~\ref{f:M10_D10} (and also online), we
show not only the time-variation of \hd\ and \mdot\ for a sample of \mdi\
triplets, but also the corresponding values of \lledd.  As discussed in
greater detail in \S\ref{s:conclusion}, the \lledd\ ratio is also helpful
in constraining the ranges of the black hole and binary parameters since
we expect BH binaries in soft state to have \lledd\ greater than a few
percent typically.

In Figs.~\ref{f:heat-5m}--\ref{f:heat-15m} we present heat maps of the
relevant fit parameters, viz. \astar, \hd, \lledd, and \nh, for \mbh=5,
10 and 15\msun\ respectively.  At lower masses we find that the best-fit
\astar\ dramatically changes from maximal retrograde spin at lower
inclinations (55\arcdeg\ and 65\arcdeg) to maximal prograde spin at higher
inclinations (75\arcdeg\ and 85\arcdeg). This rapid flip is seen for
all assumed distances (i.e. between 5--20 kpc).  Intermediate spins are
however obtained for higher black hole masses if the inclination is low.

As discussed in the previous section, \hd\ was allowed to vary between
the observations. Therefore for every \mdi\ triplet we have 26 best-fit
values of \hd. But as Fig.~\ref{f:M10_D10} (and similar figures for other
fits available online) show, the variation in \hd\ between observations
is quite small and the average value is a good indicator of the spectral
hardening for a given triplet.  We show the variation of average \hd\
in the top-right panels of Figs.~\ref{f:heat-5m}--\ref{f:heat-15m}. The
color scheme of the \hd\ heat maps is such that generally acceptable \hd\
values (\about1.5--2.5) would be green in color. Progressively higher
(and probably physically implausible) values are denoted by orange
and then red.  Irrespective of \mbh, we find that fits assuming lower
inclinations result in very large \hd, making them less likely. For higher
inclinations, we find that the region of `acceptable' \hd\ values move
progressively from 5--10 kpc for 5\msun\ to higher distances for higher
black hole masses.

In the bottom-left panel of Figs.~\ref{f:heat-5m}--\ref{f:heat-15m}
we show heat maps based on average values of Eddington fraction. As
Fig.~\ref{f:M10_D10} shows, the variations in \lledd\ from observation to
observation is larger than the variations in \hd, but still less than a
factor of \about2.  On the other hand, the average value itself changes
by \about3 orders of magnitude across the \mdi\ parameter space we have
explored. Therefore these heat maps show the general ballpark regime
where the \lledd\ values lie for any given \mdi\ triplet. As discussed
in \S\ref{s:conclusion}, this helps in constraining the ranges of the
parameter space. While these heat maps show values of \lledd\ estimated
by the {\texttt kerrbb} model, Fig.~\ref{f:heat-ledd-simple} shows the
expected range in \lledd\ if the X-ray emission is isotropic. To create
these maps we have simply assumed that the X-ray flux from the source
was $F_x$=1.5$\times$10$^{-9}$ erg s$^{-1}$ cm$^{-2}$ (the average flux
level in our observations).  Then for a given \mdi\ triplet the \lledd\
is given by
\begin{equation}
\lledd=9.5\times10^5 \frac{F_x d_{kpc}^2}{(\mbh/\msun) / cos(i)}.
\end{equation}

In the bottom-right panel of Figs.~\ref{f:heat-5m}--\ref{f:heat-15m} we
show the heat maps based on values of \nh\ that we obtain from our fits to
the different \mdi\ triplets.  These figures show that lower inclinations
prefer higher columns. Also there are some hints of decreasing column
with increasing distance.  The range of \nh\ we obtain is consistent with
independent observations made with \chandra\ and \xmm\ \citep{Nowak+2008}.

% End --- chi-sq results subsubsection /*}}}*/

\subsection{Markov Chain Monte Carlo Simulations} % /*{{{*/
\label{ss:mcmc}

While the traditional \csq\ fitting method narrowed down the region of
the parameter space with statistically better fits, the parameter space
itself was sampled quite coarsely (only at discrete \mdi\ triplets). This
is because the \csq\ fitting technique becomes extremely computationally
expensive for problems such as ours that involve a large number of
free parameters. We therefore used MCMC simulations to estimate the
best parameter values, the errors in these parameters, and to study
correlations between different parameters. As we will see below, the MCMC
validates the results from traditional \csq\ fitting, and that these
two techniques converge towards the same results further strengthens
our conclusions.

We used an in-house ISIS script that implements affine-invariant ensemble
sampler for MCMC proposed by \citet{GW2010}, in a manner similar to the
{\em emcee} python package developed by \citet{emcee}. The key advantages
of the Goodman-Weare algorithm over the commonly used Metropolis-Hastings
algorithm are that the Goodman-Weare algorithm does not require a choice
of proposal distribution, and also the Goodman-Weare algorithm can be
easily parallelized in a cluster computing environment.  We used all
the 24 CPU cores of one compute node of the \texttt{zephyr} cluster at
Wheaton College to carry out the MCMC computations.

The starting point for the MCMC is the best-fit solution obtained using
conventional \csq\ fitting for the \mdi=\{10, 10, 75\degr\} triplet.
An ensemble of model parameter sets, called walkers, are then started in
a small ball\footnote{The walkers have initial parameters distributed
as a gaussian about the best-fit value with sigma = 0.1*(max-value) or
sigma = 0.1*(value-min) (i.e., normalized to the possibly asymmetric
min/max values that a given parameter can take).  } around the
best-fit solution.  For our data set we started with 10 walkers per
free parameter.  In addition to the 54 free parameters already described
above in the section on \csq\ fitting, we also allowed the mass of the
accretor, distance to the system, and the inclination to vary in the MCMC
simulations. Thus the total number of walkers in our simulations were 540.
The MCMC simulations explored the following parameter ranges:
$i\Rightarrow[5\degr, 85\degr]$,
$\mdot\Rightarrow[10^{-4}, 20]\times10^{18}$ $g/s$,
$\astar\Rightarrow[-1, 0.9999]$,
$\mbh\Rightarrow[2, 35]$ $\msun$,
$d\Rightarrow[3, 30]$ kpc,
$\hd\Rightarrow[1,10]$, and
$\nh\Rightarrow[0.01,1.0]\times10^{22}$ cm$^{-2}$.
% End --- MCMC setup /*}}}*/

\subsubsection{Results of MCMC simulations} % Results of MCMC/*{{{*/
It required about 27,000 steps for the ensemble of walkers in the
simulation to attain equillibrium. Data generated during the initial
stages were excluded from final analysis. Here we present results from
a chain of 5,757,000 elements after rejecting data from the initial
burn-in period.

Fig.~\ref{f:parcor} shows the probability density functions for the
column density, inclination, and spin parameter. The marginalized 1D
histograms along the diagonal clearly show a peaked distribution for
\nh\ and $i$. The constraint on the inclination is the most stringent
constraint on the inclination of the inner accretion disk for this
system so far. The constraint on column density is consistent with that
measured previously from high-resolution X-ray grating observations
from {\em Chandra} and {\em XMM-Newton} \citep{Nowak+2008}.  Also,
the MCMC derived spin parameter is consistent with a maximally spinning
prograde black hole, and we only obtain a lower limit on the value of
the spin parameter.  The minimum-\csq\ model from the MCMC simulations
has a \csq\ value of 15197.9. The residuals for all 26 observations for
this model are shown in Figs.~\ref{f:mcmc_res1}--\ref{f:mcmc_res2}.
Tables~\ref{tab:mcmcglobalres} and~\ref{tab:mcmclocalres} list the
minimum-\csq\ model values and 90\% confidence intervals for the global
(i.e. \nh, $i$, \astar) and local (i.e.  \mdot$_{disk}$ and \hd\
for each observation) model parameters based on the MCMC simulations.
As in the case of \csq\ fitting, no constraint could be obtained for
the accretor's mass or the distance to the system.

The off-diagonal contour plots in Fig.~\ref{f:parcor} show the correlation
between different model parameters.  The correlation between these
parameters can be qualitatively explained via the following line of
reasoning: the more face on the disk is, the more we see its inner
regions $\Rightarrow$ the spectrum will be more strongly affected by
gravitational redshift and appear softer $\Rightarrow$ the more \nh\
we need to keep the same spectrum, or the more spin to boost it up to
slightly higher temperature.

Overall, the MCMC results are completely consistent with the \csq\ fitting
results (but the MCMC technique, being better suited to address problems
with large number of free parameters than \csq\ fitting, gives more
precise results).  The convergence of the two techniques give additional
confidence not only about the implementation of the techniques but also
our the conclusions.

% End --- results of MCMC /*}}}*/

% End - analysis section /*}}}*/

\section{Discussion and Conclusions} \label{s:conclusion} % /*{{{*/

We have presented uniform spectral analysis of all X-ray data of
\src\ taken by \swift's \textsl{XRT}, where we analyze the data using
traditional \csq\ fitting technique as well as using MCMC simulations.
While present computational resources make it prohibitively expensive
to explore a finer grid of the parameter space using traditional \csq\
technique, the MCMC simulations validate the trends suggested by \csq\
fitting and allow us to explore the parameters much more precisely and
also study the correlation between various model parameters.

Both techniques point toward a relatively high-inclination of the
inner accretion disk. The \csq\ fitting, which could be done only on a
coarse grid of \mdi\ triplets due to computational limits, prefers an
inclination of \about75\degr.  Not only are the \csq\ values smallest
at $i$\about75\degr\ (among the grid points separated by 10\degr\
in inclination), but the spectral hardening factor also lies in an
`acceptable' range (\about1.5--2.5) for $i$\about75\degr.

The MCMC simulations, which prefer $i$=\inclval\ degrees, put the most
stringent constraint on the inclination of the X-ray emitting disk so
far. This is consistent with the previous upper limit of \about85\degr\
based on the absence of any X-ray eclipse. This inclination can also
be considered marginally consistent with the optical results where
\citet{Mason+2012} and \citet{Bayless+2011} concluded that the orbital
inclination was ``nearly unconstrained with permitted inclinations of
\about20\degr$<i<$70\degr''. We would however like to point out that the
inclination we measure in this work is that of the very innermost parts
of the accretion disk, closest to that of the black hole.  Since the
spin very high, we expect that this inclination is also what the black
hole's spin axis makes to our line of sight due to Bardeen-Petterson
\citeyearpar{BP1975} effect.  If the direction of the black hole's spin
angular momentum vector is different from that of the angular momentum
vector of the binary orbit (which is measured from optical observations),
that could explain any discrepancy between the inclination measured in
X-rays and optical.

\begin{figure} % Heat maps for 5M /*{{{*/
 \includegraphics[height=0.49\textwidth, angle=-90]{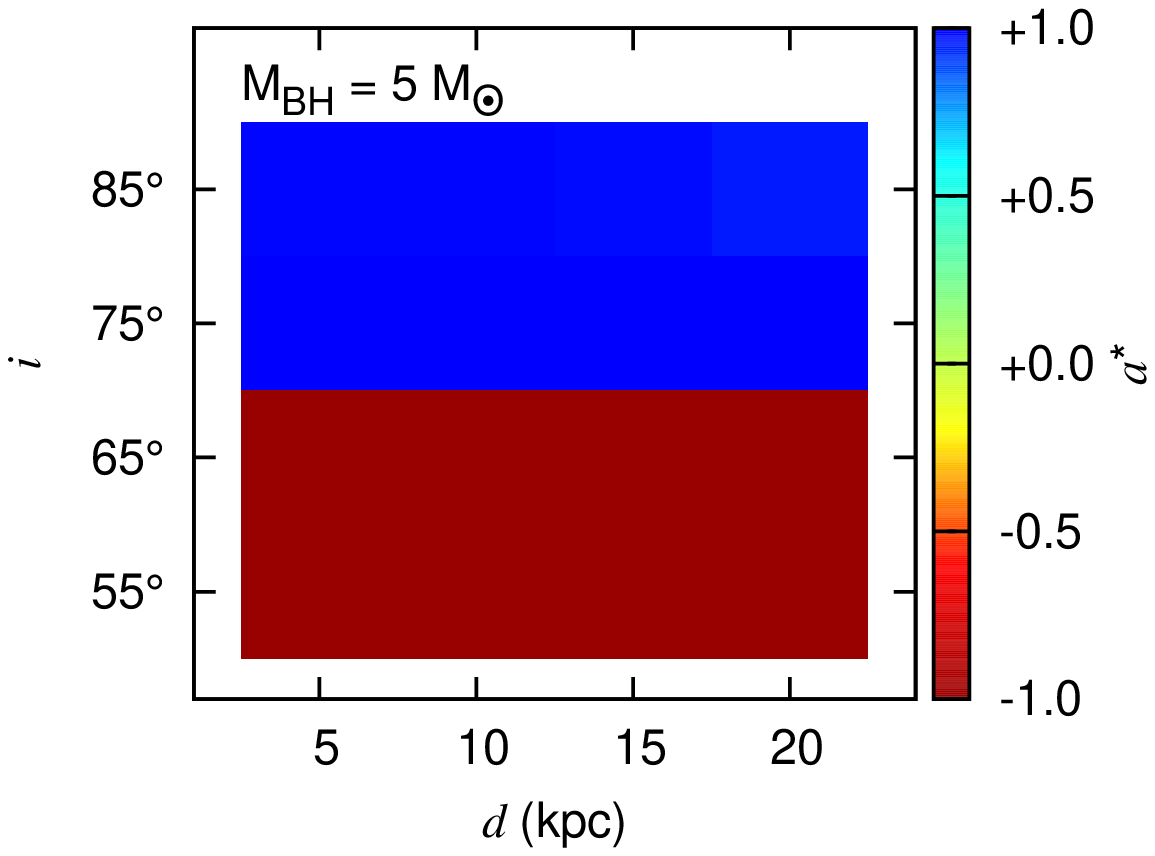}
 \includegraphics[height=0.49\textwidth, angle=-90]{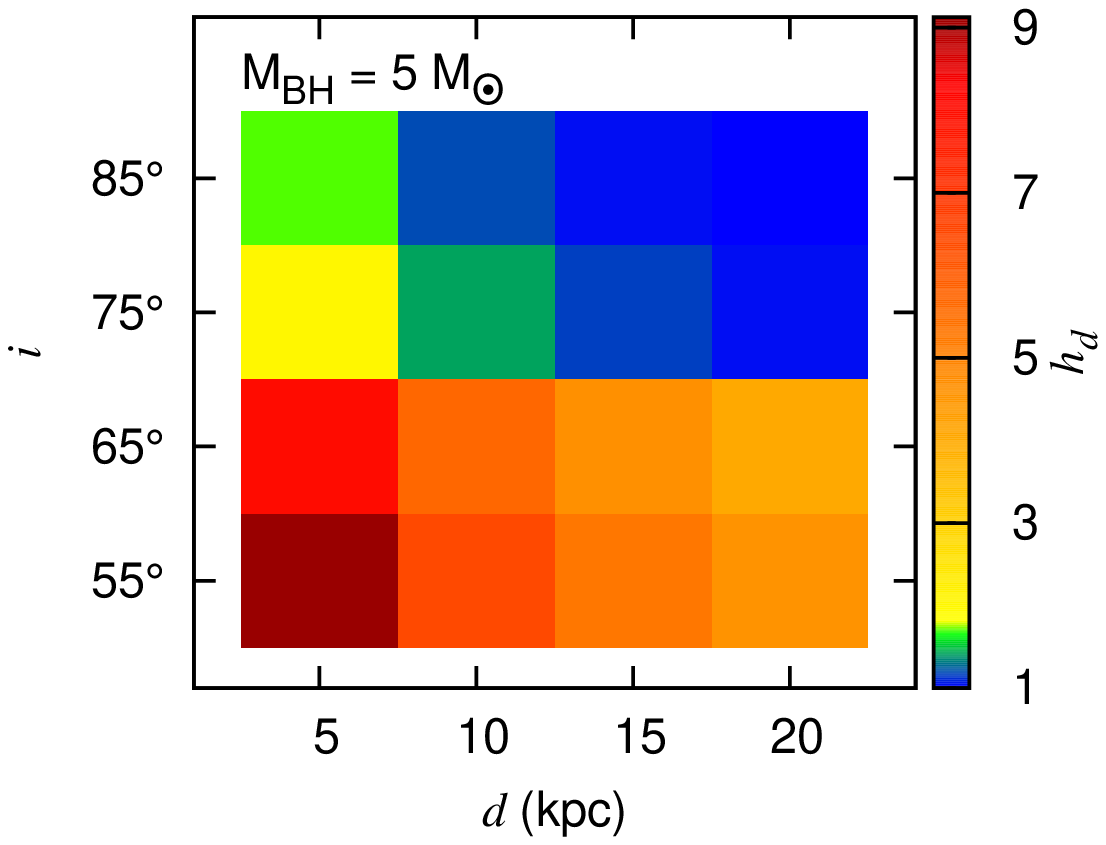}
 \includegraphics[height=0.49\textwidth, angle=-90]{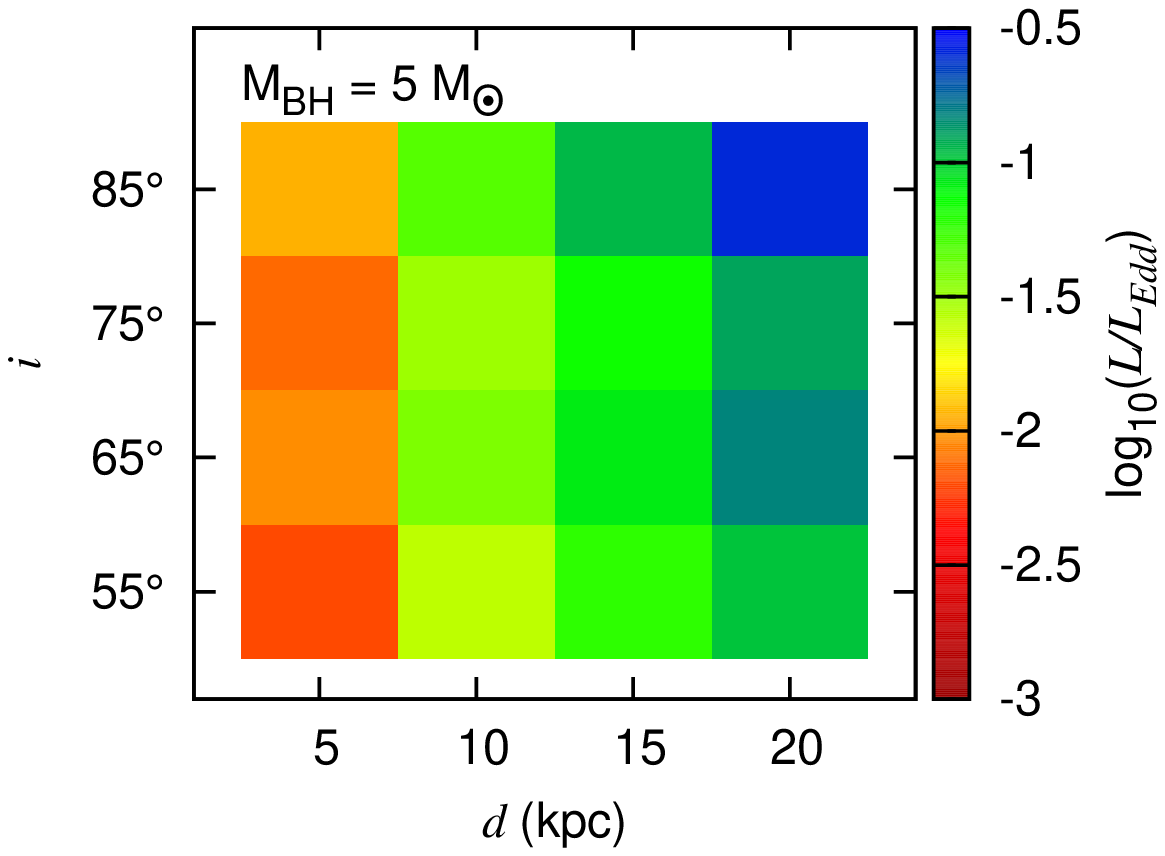}
 \includegraphics[height=0.49\textwidth, angle=-90]{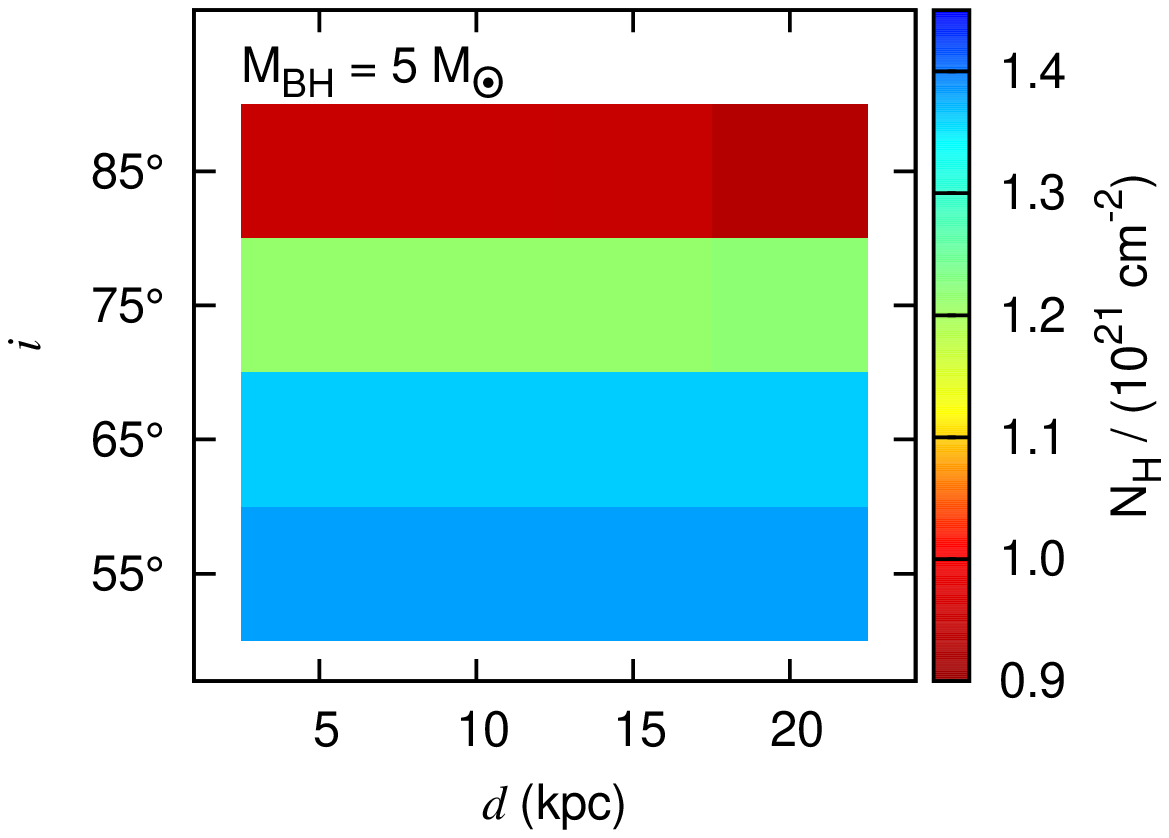}
 \caption{Heat maps based on values of best-fit parameters, assuming
 \mbh=5\msun.
 The top-left, top-right, bottom-left, and bottom-right panels are for 
 \astar, \hd, \lledd, and \nh\ respectively.
 For the top-left (\astar) and bottom-right (\nh) panels each \mdi\ triplet 
 is represented by a square whose color is indicative of the 
 best-fit parameter value obtained by simultaneously fitting all 26
 observations.
 For the top-right (\hd) and bottom-left (\lledd) panels each \mdi\ triplet
 is represented by a square whose color is indicative of the 
 average of 26 best-fit values obtained by simultaneously fitting all 26
 observations. Note the following:
 \newline {\bf \astar} -- The best-fit value of the spin parameter
 rapidly changes between lower inclinations (55\arcdeg, 65\arcdeg) and 
 higher inclinations (75\arcdeg, 85\arcdeg). See text, especially 
 \S\ref{s:conclusion}, for details.
 \newline {\bf \hd} --  Values of \hd\ in the range of
 $\sim$1.5-2.5 are physically plausible, and indicated by green and
 yellow color.
 Low inclination models ($i=55\arcdeg, 65\arcdeg$) are strongly disfavored 
 because of the unphysically high values of \hd. For higher inclinations, the 
 regions of `acceptable' \hd\ values move progressively from 5--10 kpc 
 for $\sim$5\msun to higher distances for higher accretor masses.
 \newline {\bf \lledd} -- The typical \lledd\ luminosity range spanned by 
 X-ray binaries in soft state is encompassed by yellow,
 green, and bluish colors in our color scheme for this figure.
 Luminosities close to Eddington would drive strong winds that are not 
 seen in this system. Therefore high \lledd\ are unlikely. Smaller 
 distances ($\sim$5 kpc) are not favored by any choices of the mass because
 the best-fit luminosities are too low.  Higher accretor masses would 
 require higher distances for the luminosity to be in the comfort zone for
 a soft state X-ray binary. 
 \newline {\bf \nh} -- Lower inclinations prefer higher columns.
 The range of \nh\ values is consistent with independent observations 
 made with other instruments.
 }
 \label{f:heat-5m} 
\end{figure}

\begin{figure} % Heat maps for 10M /*{{{*/
 \includegraphics[height=0.49\textwidth, angle=-90]{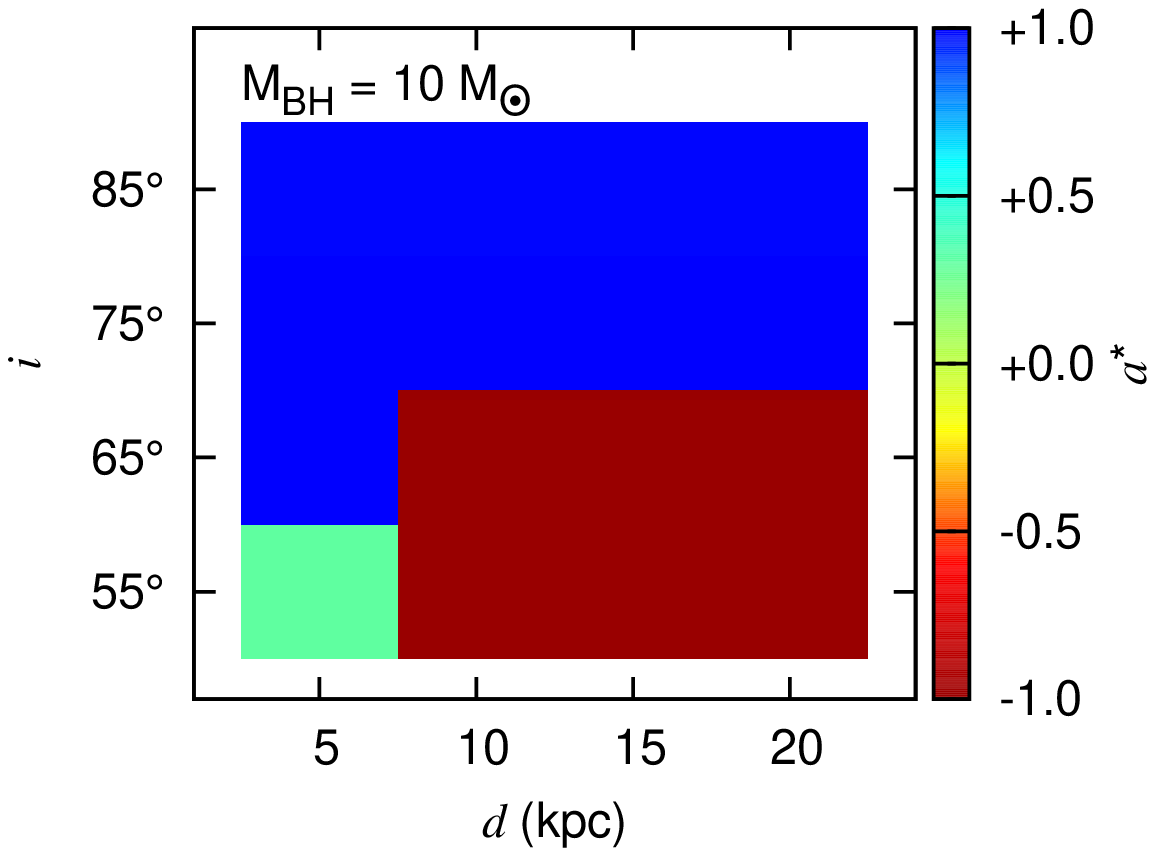}
 \includegraphics[height=0.49\textwidth, angle=-90]{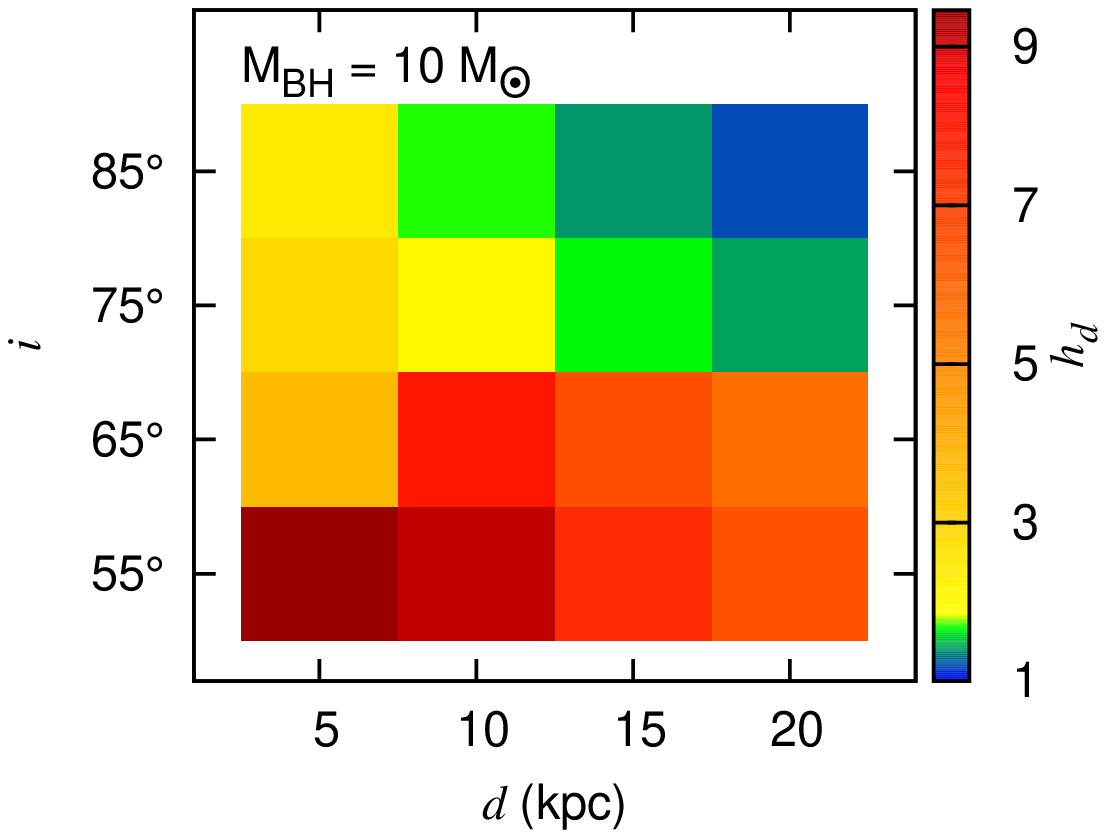}
 \includegraphics[height=0.49\textwidth, angle=-90]{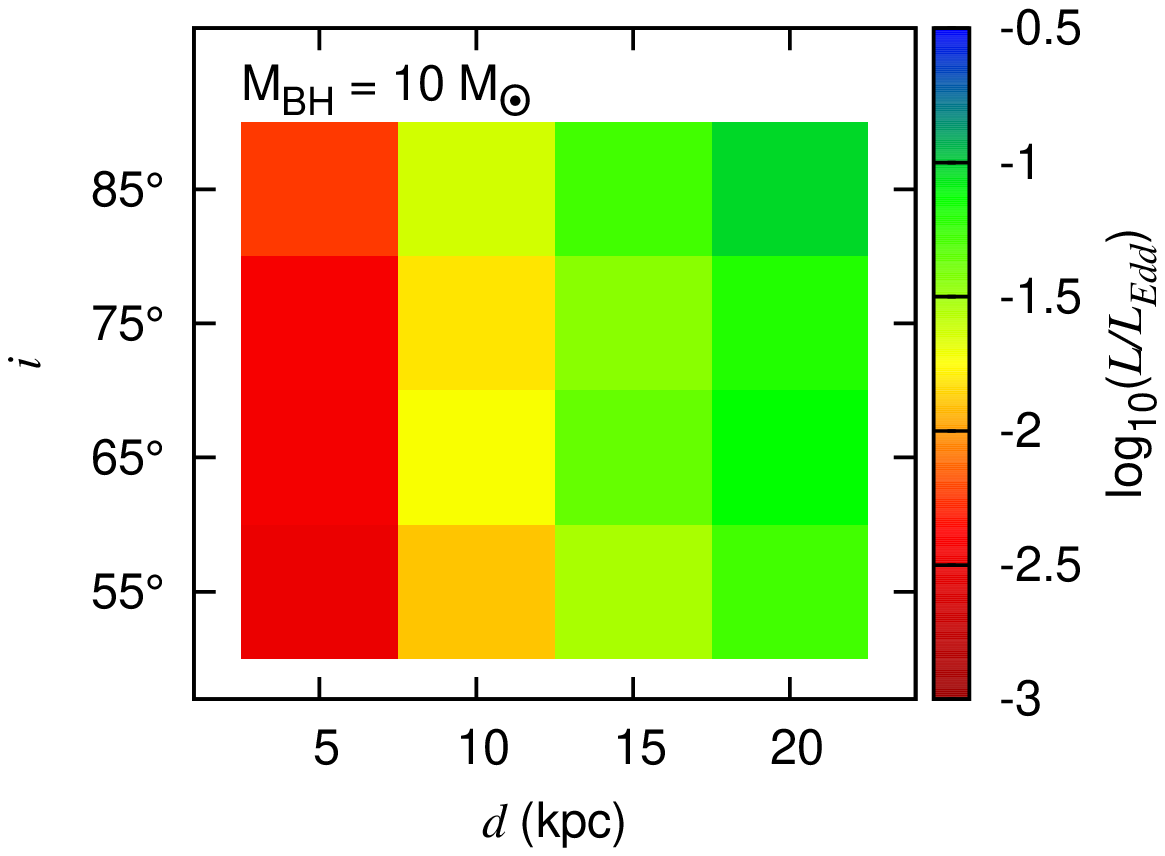}
 \includegraphics[height=0.49\textwidth, angle=-90]{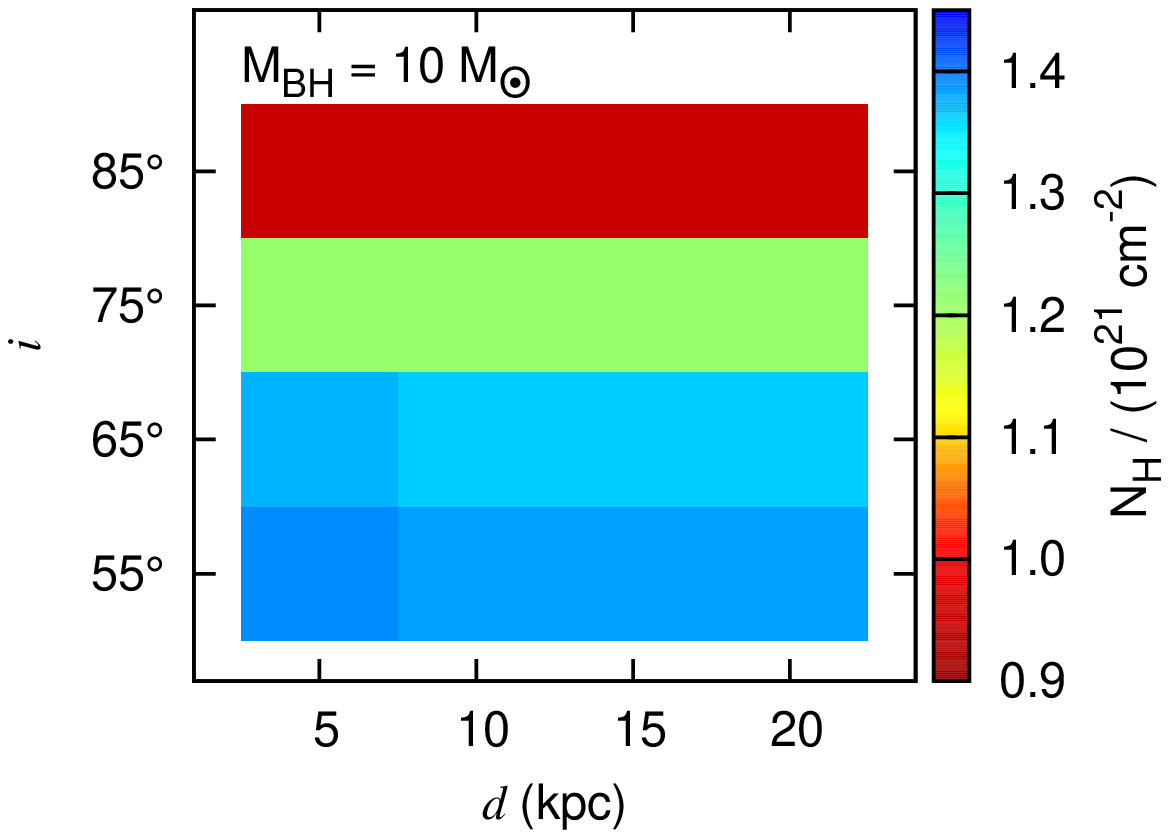}
 \caption{Same as Fig.~\ref{f:heat-5m}, but for \mbh=10\msun.
 }
 \label{f:heat-10m} 
\end{figure}

\begin{figure} % Heat maps for 15M /*{{{*/
 \includegraphics[height=0.49\textwidth, angle=-90]{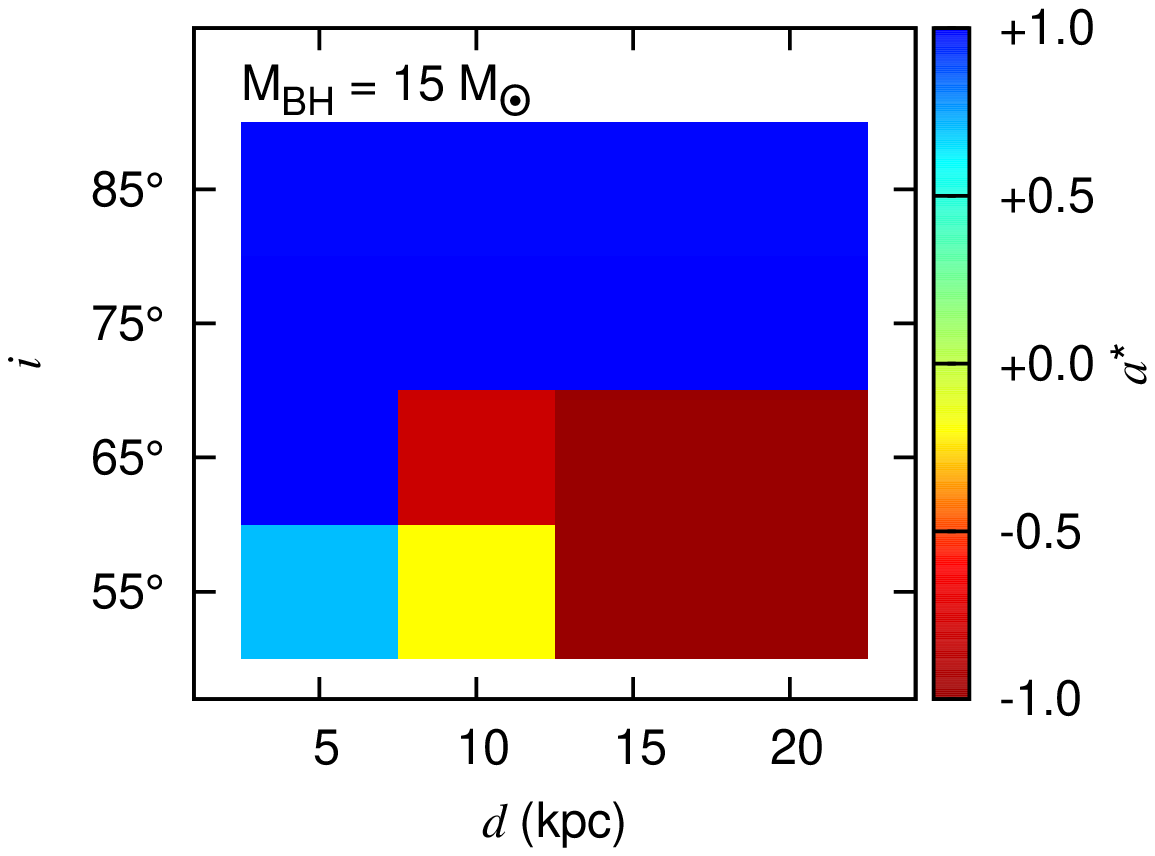}
 \includegraphics[height=0.49\textwidth, angle=-90]{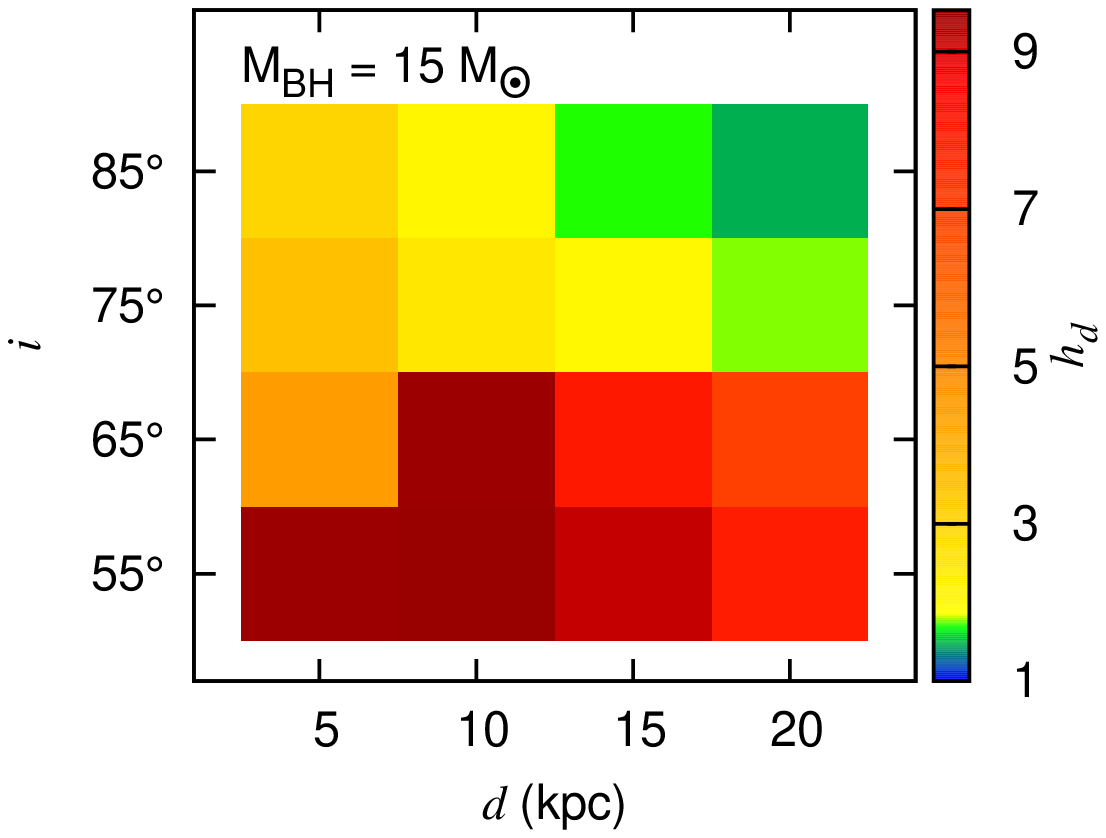}
 \includegraphics[height=0.49\textwidth, angle=-90]{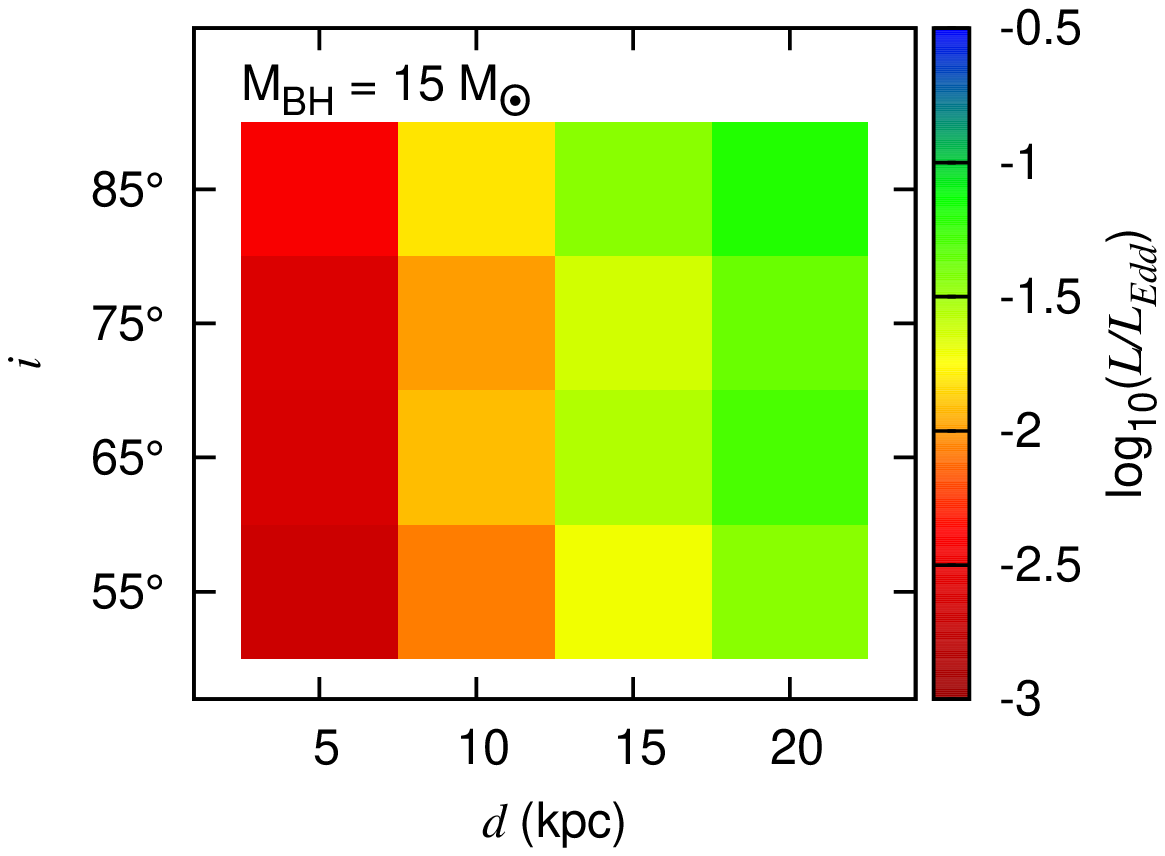}
 \includegraphics[height=0.49\textwidth, angle=-90]{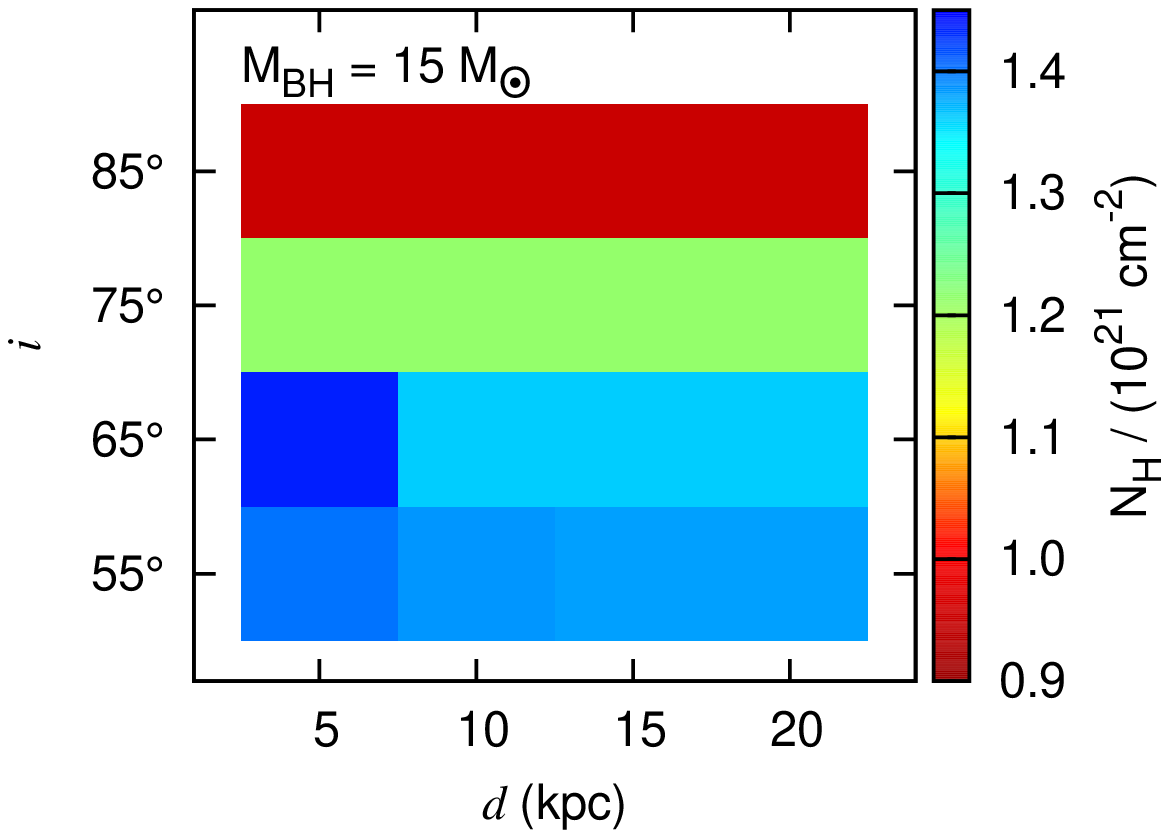}
 \caption{Same as Fig.~\ref{f:heat-5m}, but for \mbh=15\msun.
 }
 \label{f:heat-15m} 
\end{figure}

\begin{figure} % Heat maps based on simple, isotropic L_Edd /*{{{*/
 \heatmapsimplelledd{0.33}{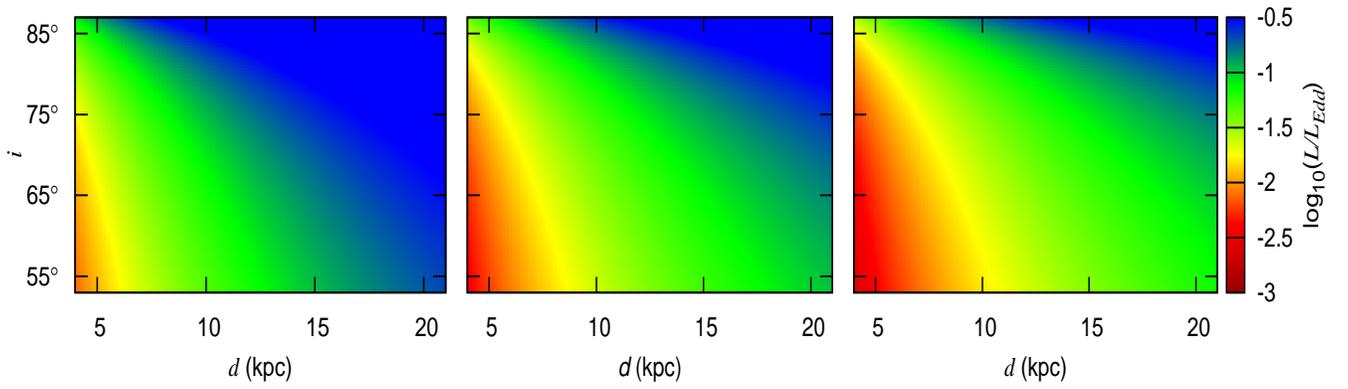}
 \caption{Heat maps based on computing \lledd\ assuming
 an isotropic source flux of 1.5$\times$10$^{-9}$ erg s$^{-1}$ 
 cm$^{-2}$. The left, middle and right panels are for 
 $M$ = 5, 10, and 15\msun\ respectively.
 }
 \label{f:heat-ledd-simple} 
\end{figure}

\begin{figure} % MCMC bestfit residuals /*{{{*/
\centering
 \begin{tabular}{c c c }
 {\centering
 \includegraphics[width=52mm, height=21mm, trim = 10mm 28mm 3mm 0mm, clip]{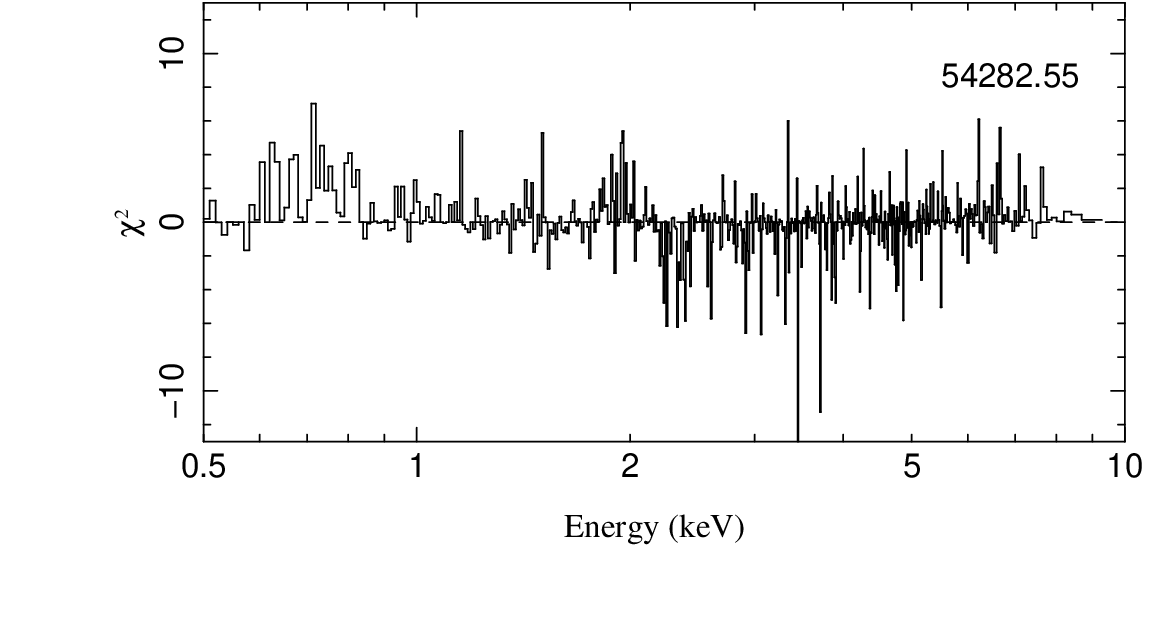}
 } & 
 {\centering
 \includegraphics[width=52mm, height=21mm, trim = 32mm 28mm 3mm 0mm, clip]{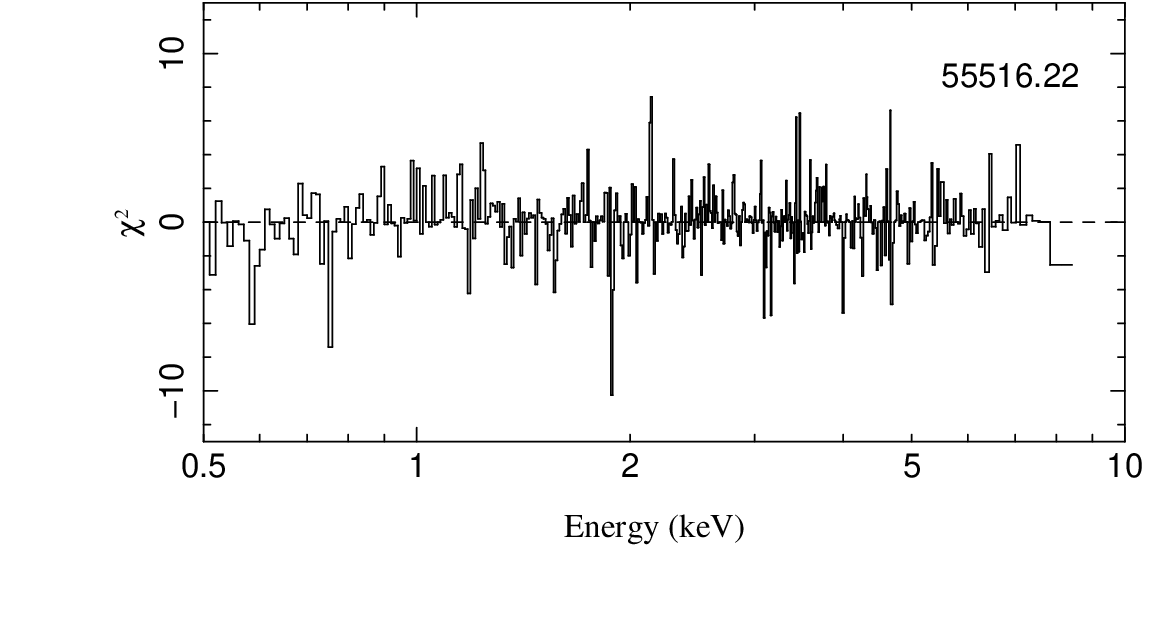}
 } & 
 {\centering
 \includegraphics[width=52mm, height=21mm, trim = 32mm 28mm 3mm 0mm, clip]{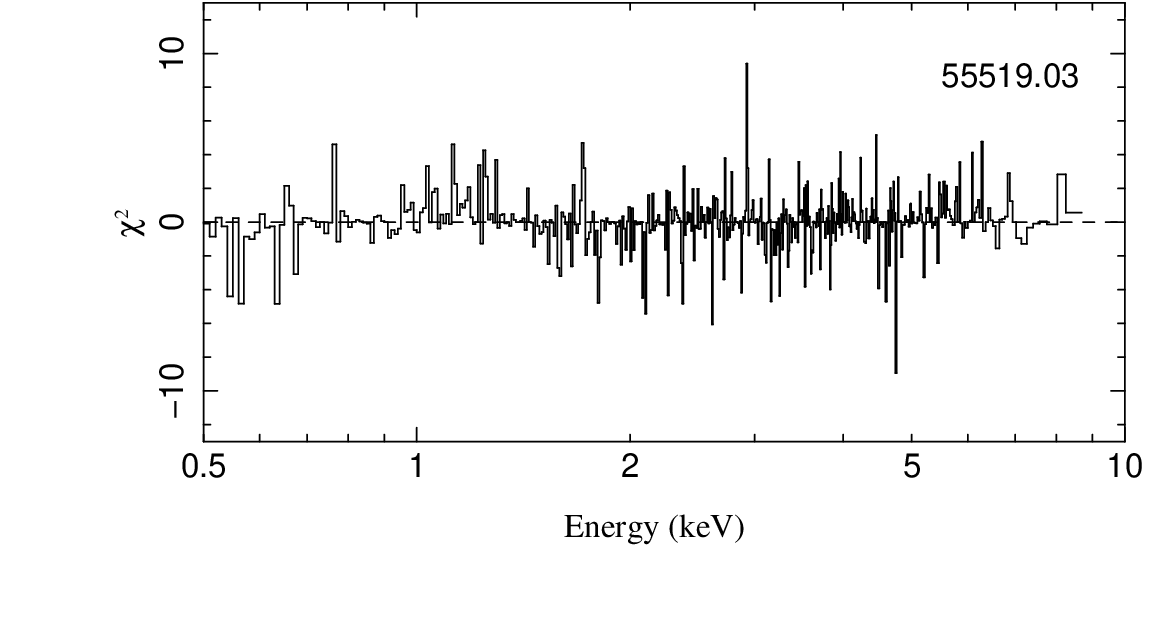}
 } \\
 {\centering
 \includegraphics[width=52mm, height=21mm, trim = 10mm 28mm 3mm 0mm, clip]{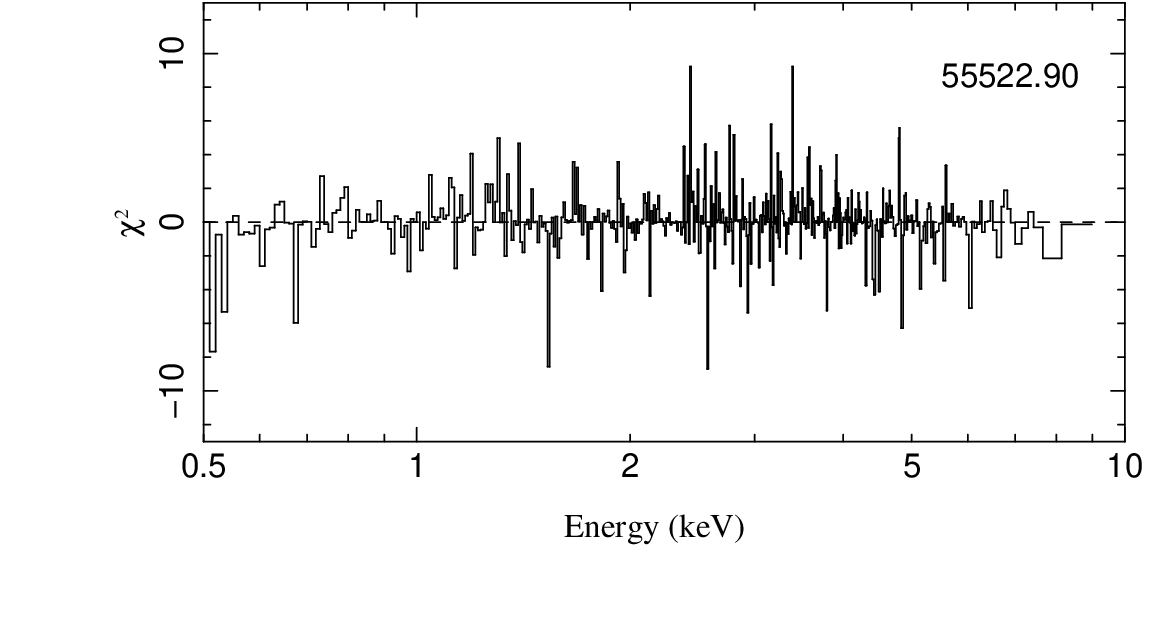}
 } & 
 {\centering
 \includegraphics[width=52mm, height=21mm, trim = 32mm 28mm 3mm 0mm, clip]{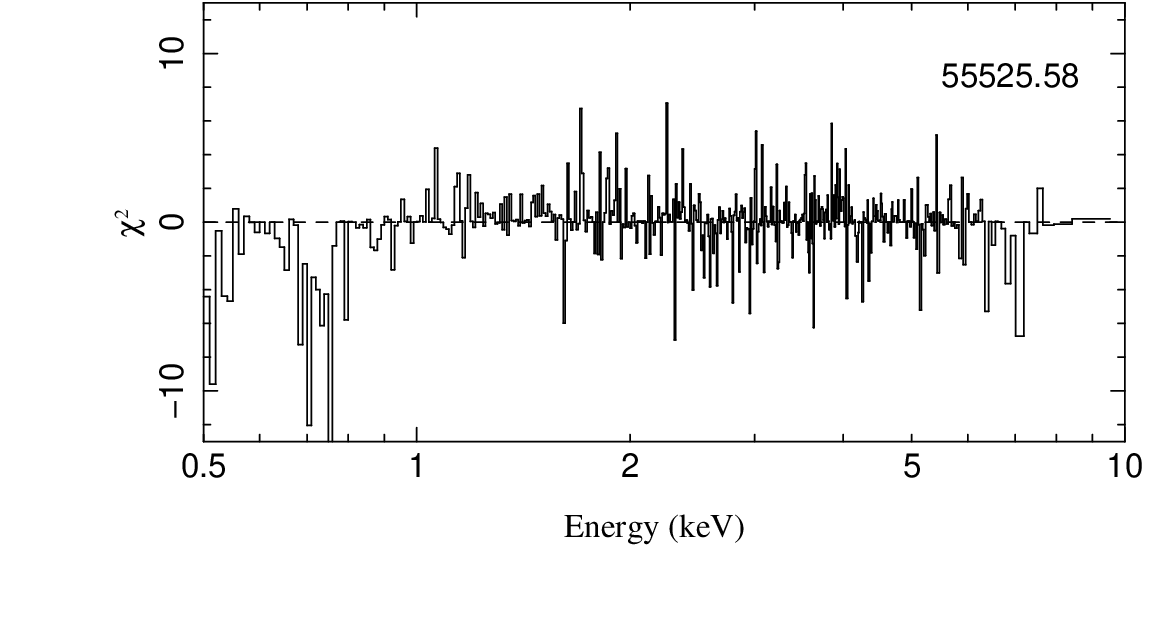}
 } & 
 {\centering
 \includegraphics[width=52mm, height=21mm, trim = 32mm 28mm 3mm 0mm, clip]{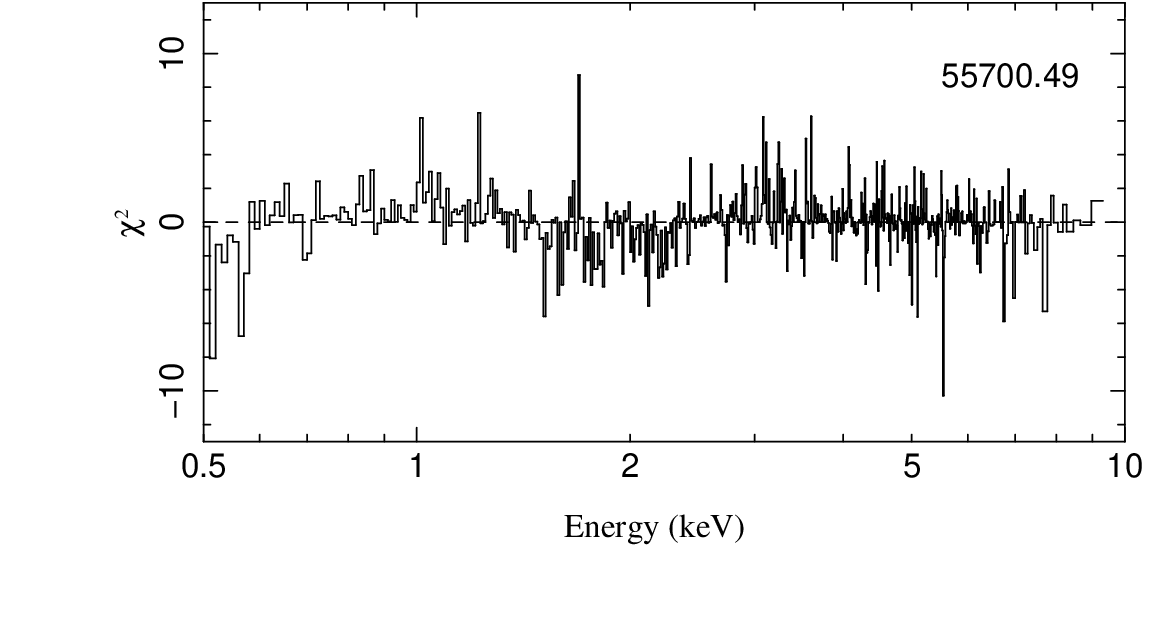}
 } \\
 {\centering
 \includegraphics[width=52mm, height=21mm, trim = 10mm 28mm 3mm 0mm, clip]{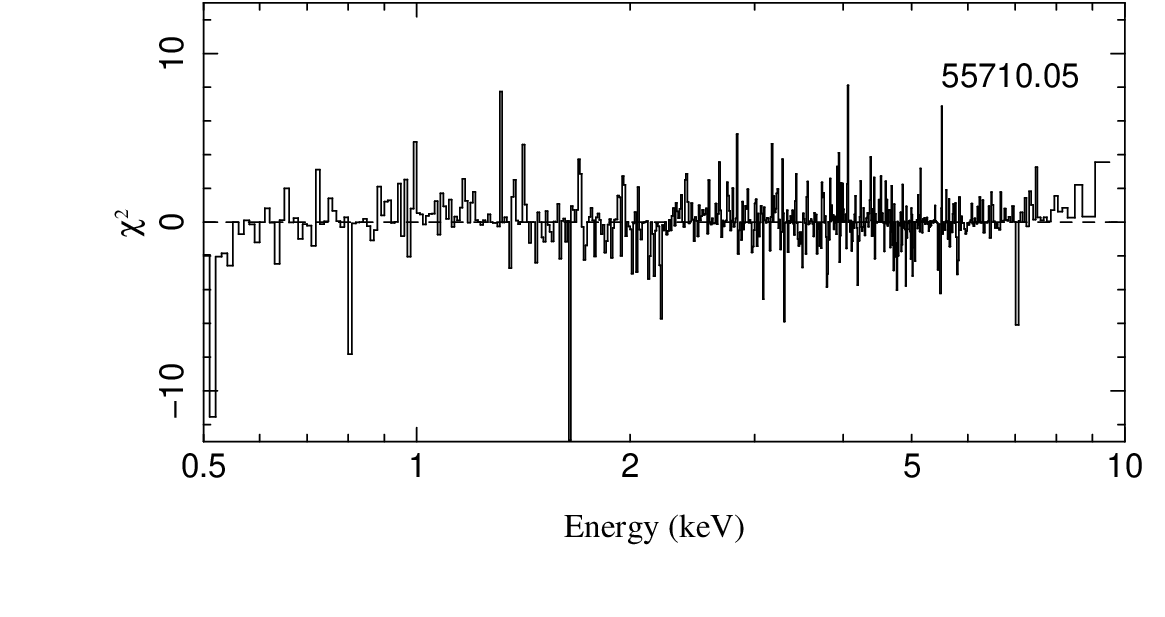}
 } & 
 {\centering
 \includegraphics[width=52mm, height=21mm, trim = 32mm 28mm 3mm 0mm, clip]{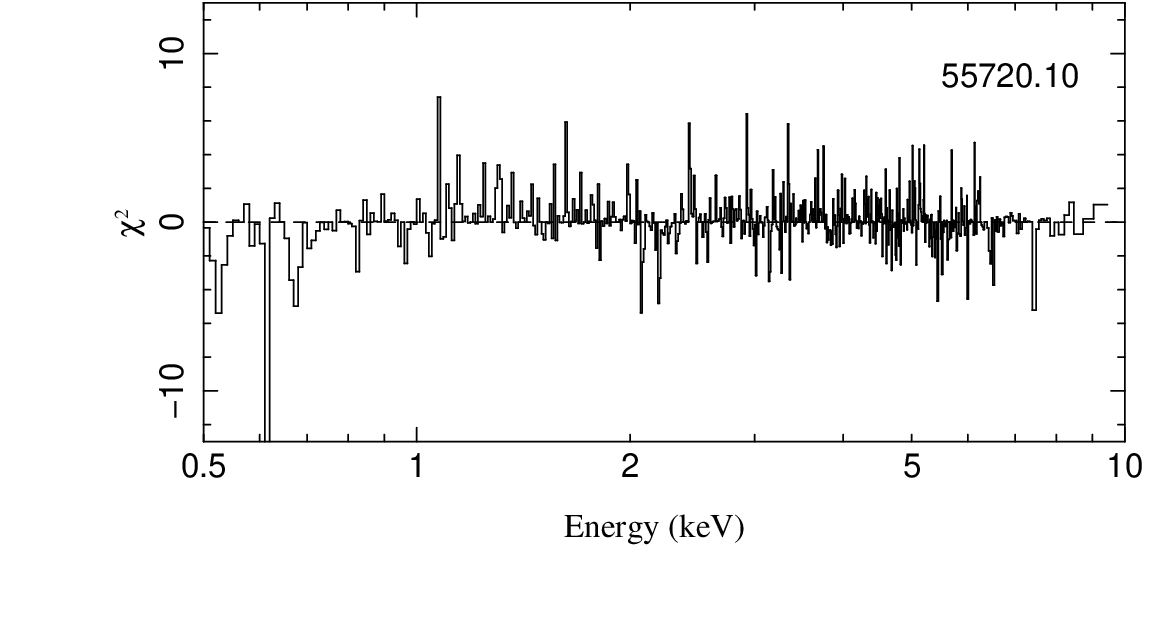}
 } & 
 {\centering
 \includegraphics[width=52mm, height=21mm, trim = 32mm 28mm 3mm 0mm, clip]{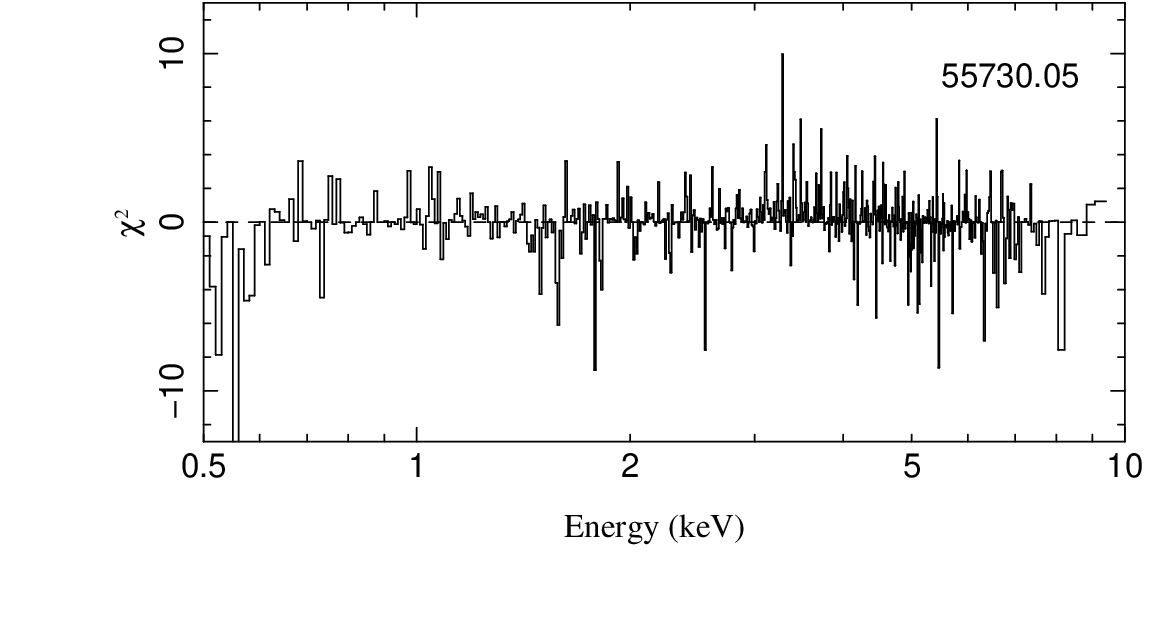}
 } \\
 {\centering
 \includegraphics[width=52mm, height=21mm, trim = 10mm 28mm 3mm 0mm, clip]{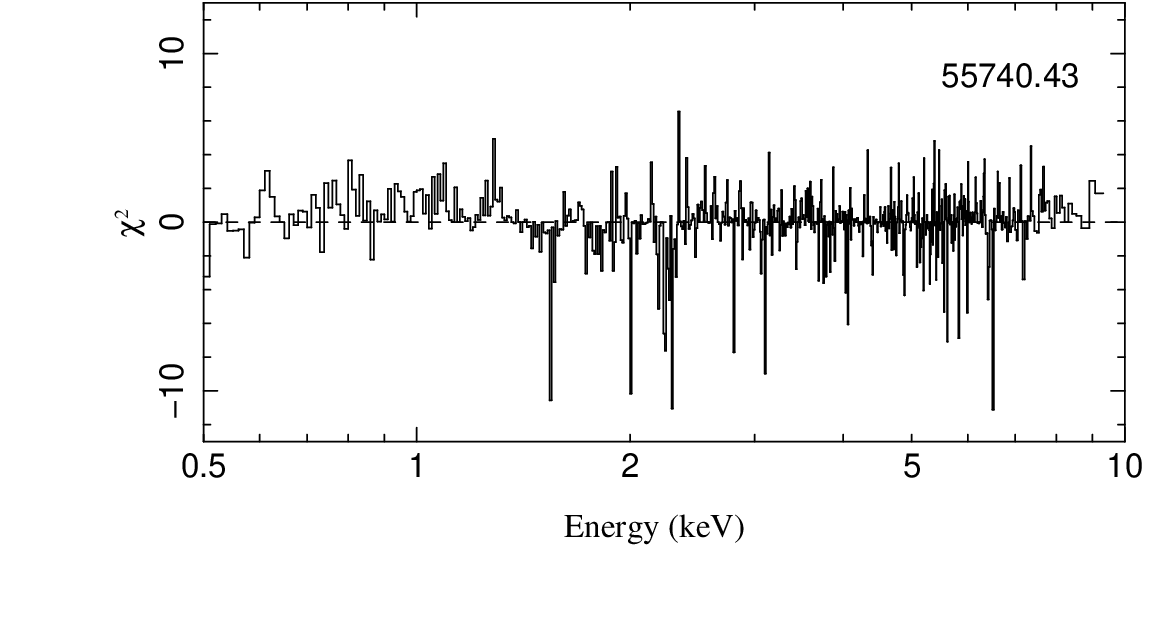}
 } & 
 {\centering
 \includegraphics[width=52mm, height=21mm, trim = 32mm 28mm 3mm 0mm, clip]{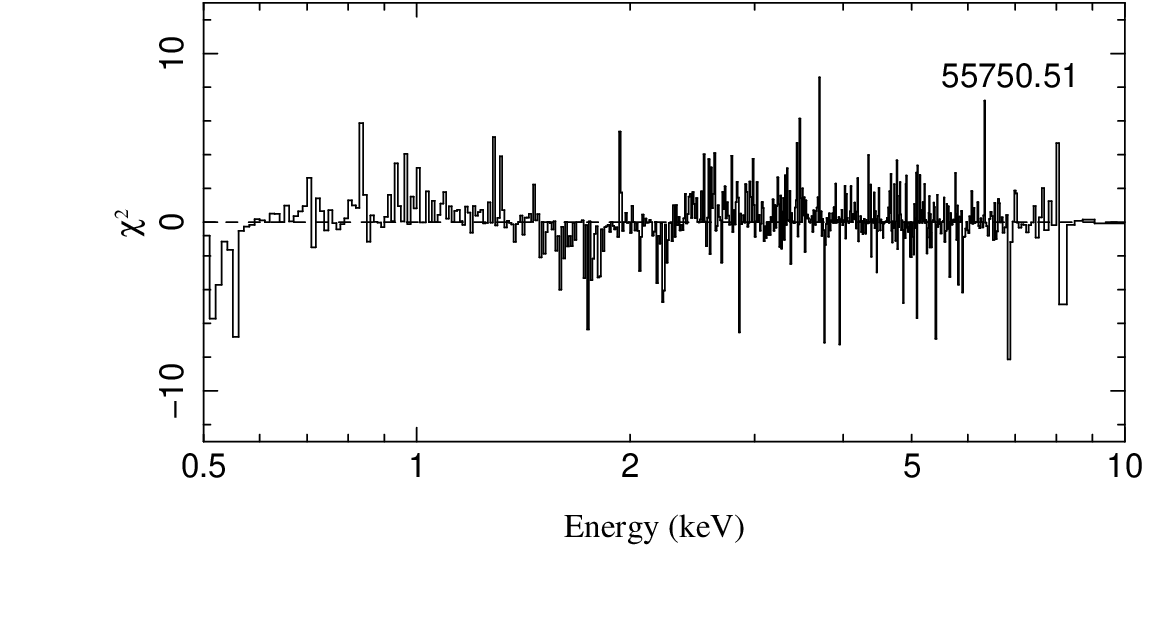}
 } & 
 {\centering
 \includegraphics[width=52mm, height=21mm, trim = 32mm 28mm 3mm 0mm, clip]{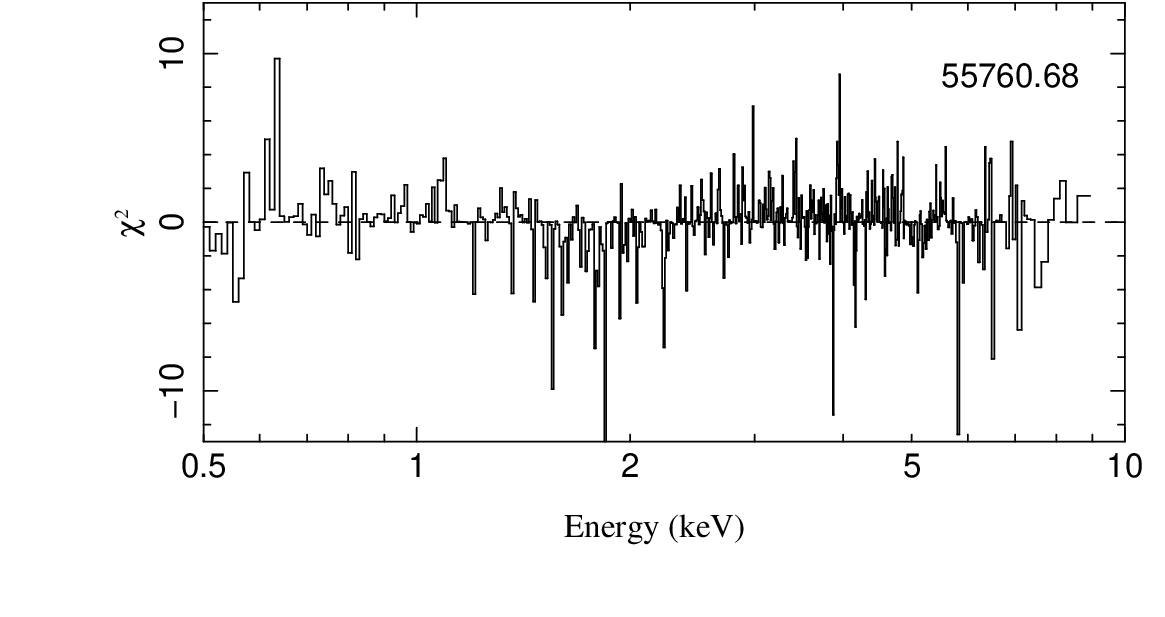}
 } \\
 {\centering
 \includegraphics[width=52mm, height=21mm, trim = 10mm 28mm 3mm 0mm, clip]{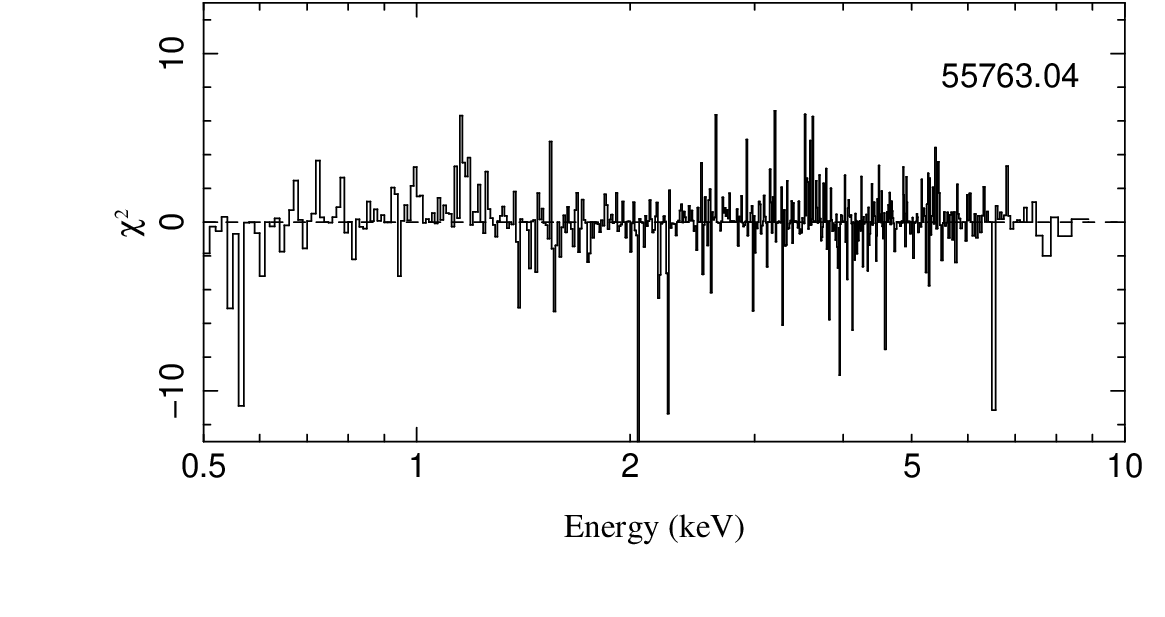}
 } & 
 {\centering
 \includegraphics[width=52mm, height=21mm, trim = 32mm 28mm 3mm 0mm, clip]{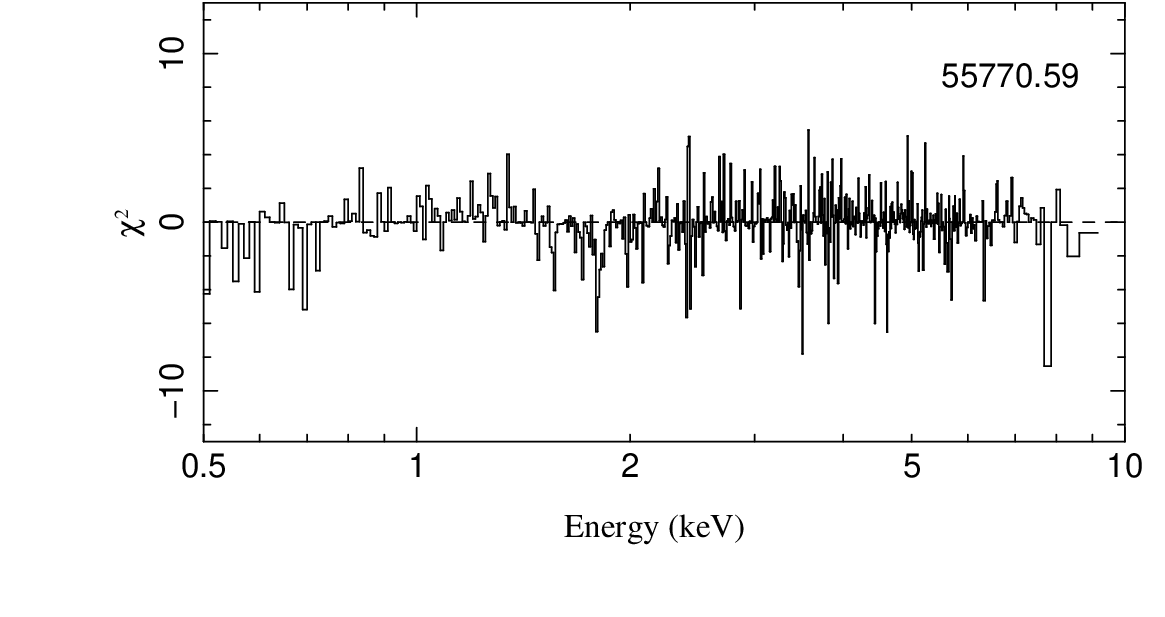}
 } & 
 {\centering
 \includegraphics[width=52mm, height=21mm, trim = 32mm 28mm 3mm 0mm, clip]{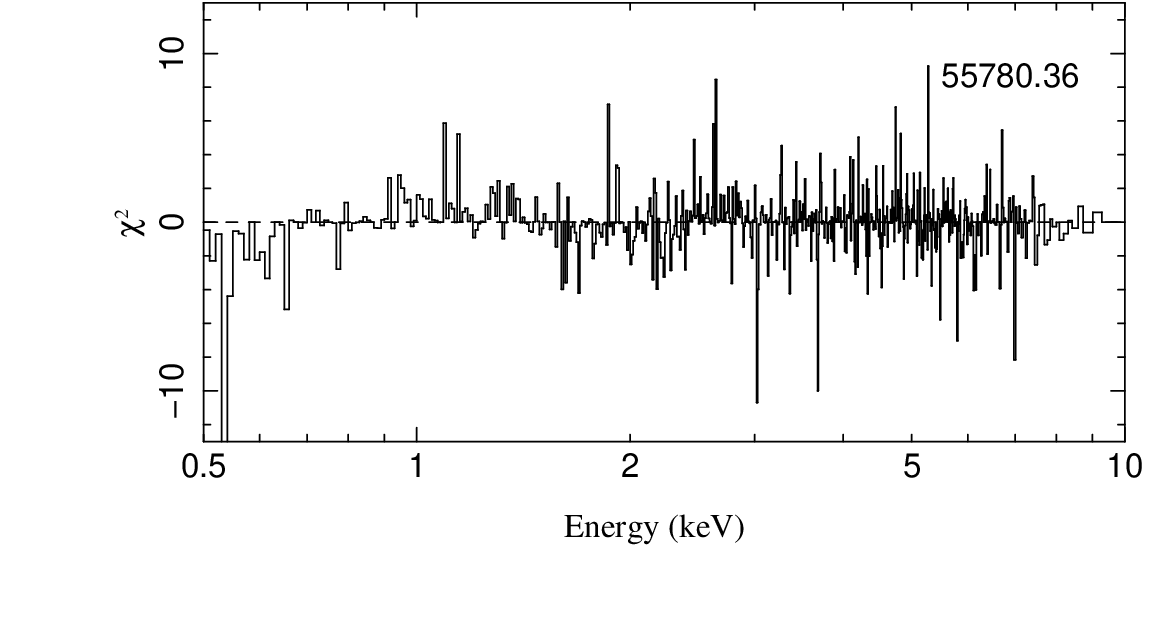}
 } \\
 {\centering
 \includegraphics[width=52mm, height=21mm, trim = 10mm 28mm 3mm 0mm, clip]{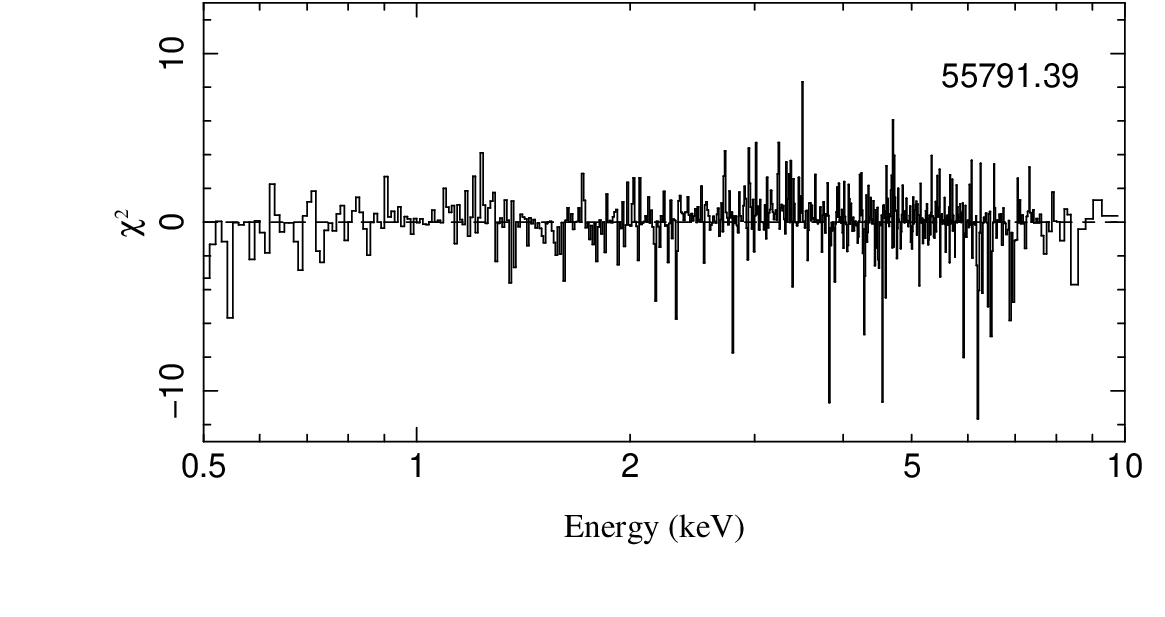}
 } & 
 {\centering
 \includegraphics[width=52mm, height=21mm, trim = 32mm 28mm 3mm 0mm, clip]{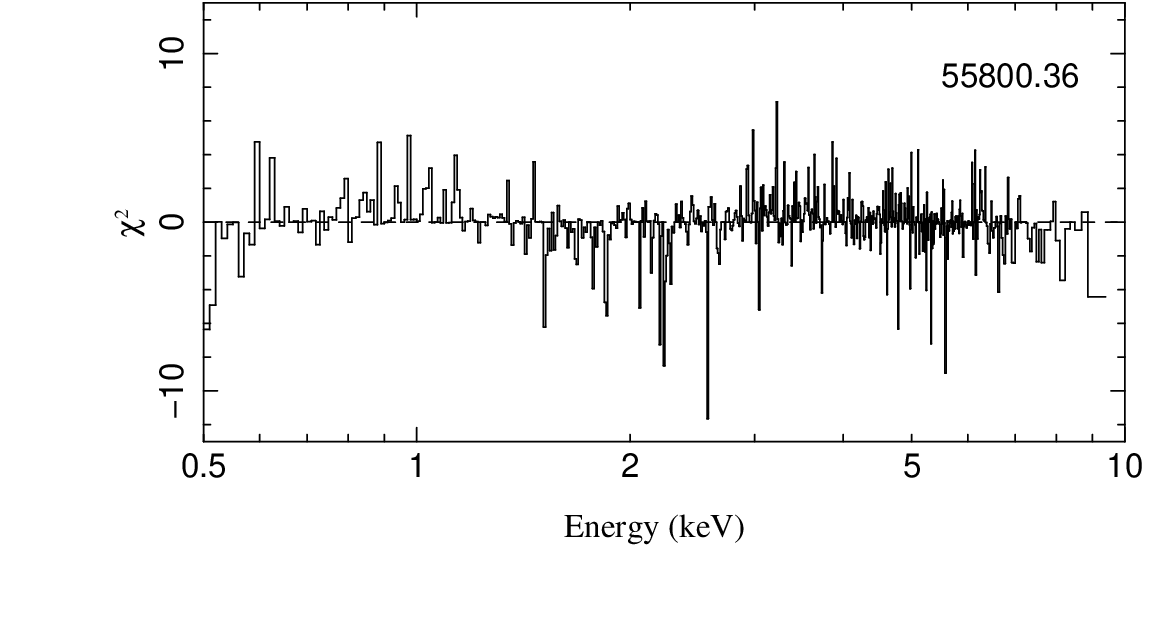}
 } & 
 {\centering
 \includegraphics[width=52mm, height=21mm, trim = 32mm 28mm 3mm 0mm, clip]{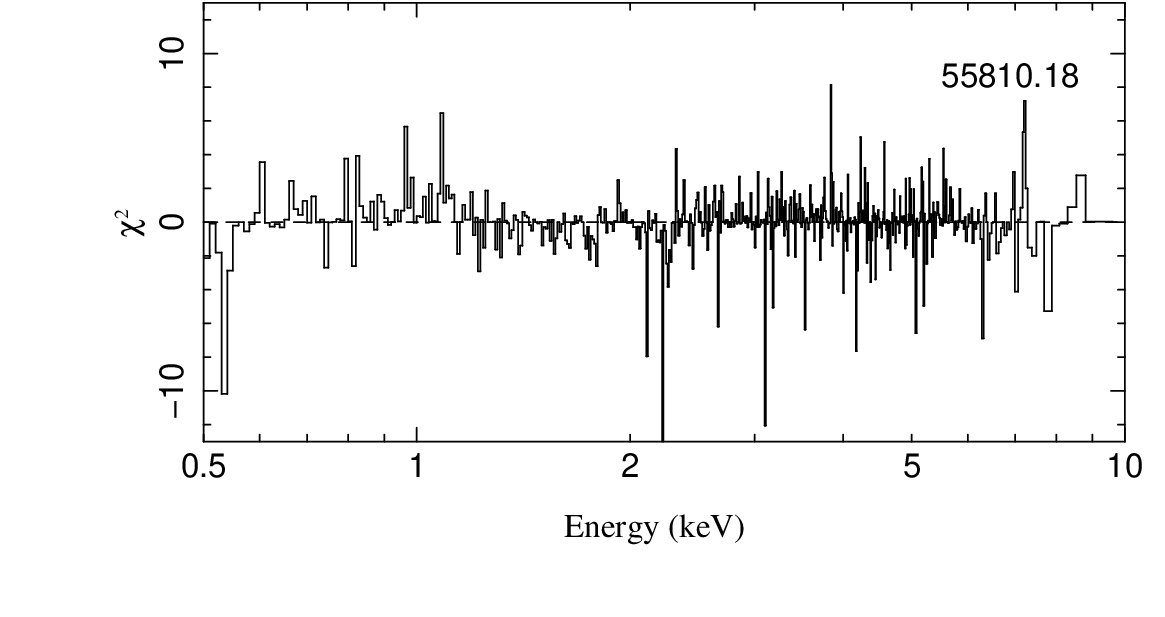}
 } \\
 {\centering
 \includegraphics[width=52mm, height=21mm, trim = 10mm 28mm 3mm 0mm, clip]{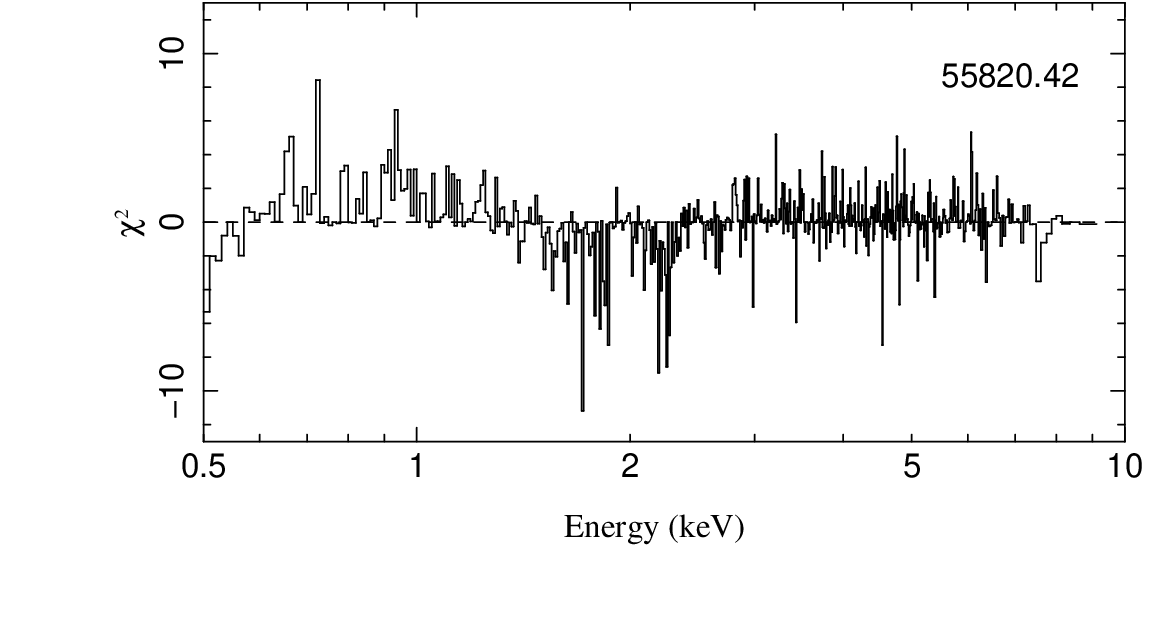}
 } &
 {\centering
 \includegraphics[width=52mm, height=21mm, trim = 32mm 28mm 3mm 0mm, clip]{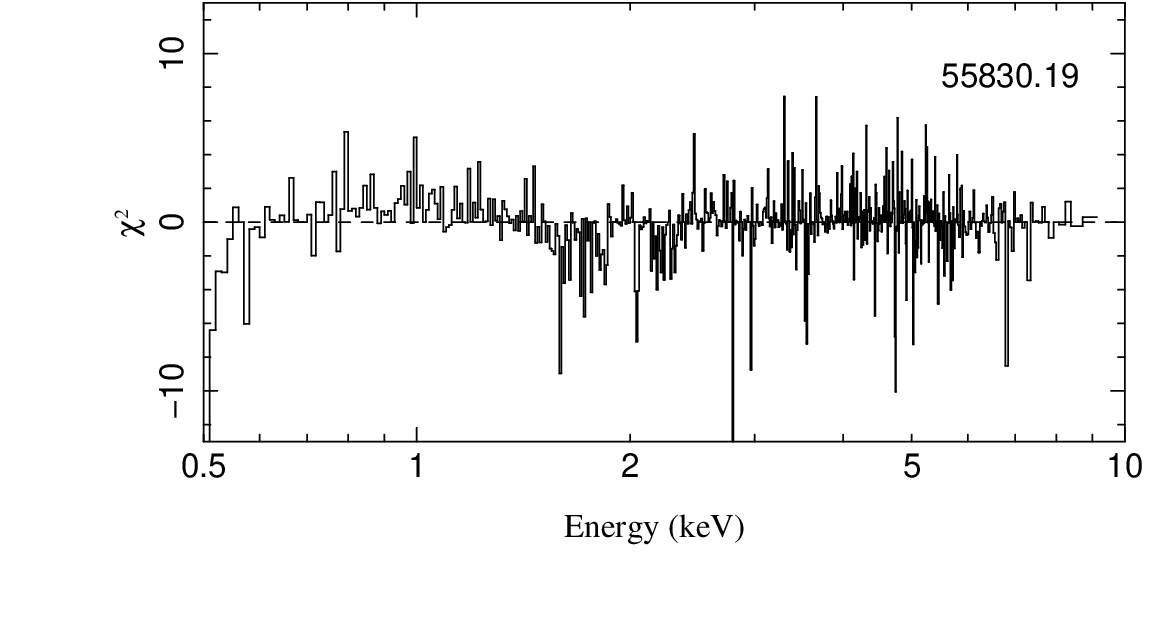}
 } &
 {\centering
 \includegraphics[width=52mm, height=21mm, trim = 32mm 28mm 3mm 0mm, clip]{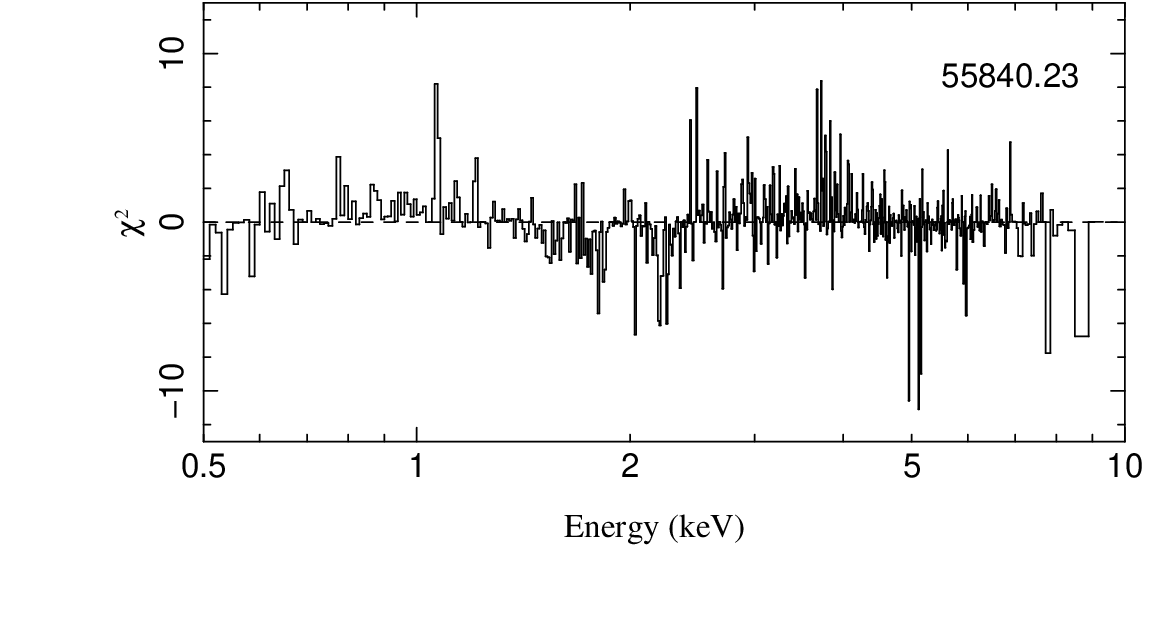}
 } \\
 {\centering
 \includegraphics[width=52mm, height=21mm, trim = 10mm 08mm 3mm 0mm, clip]{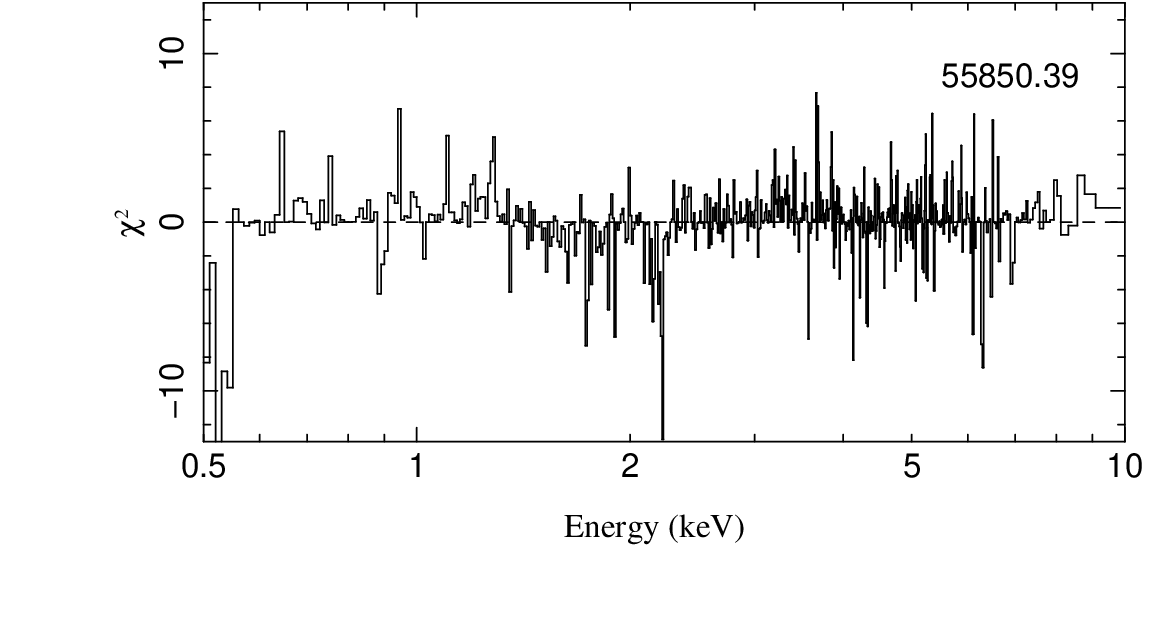}
 } &
 {\centering
 \includegraphics[width=52mm, height=21mm, trim = 32mm 08mm 3mm 0mm, clip]{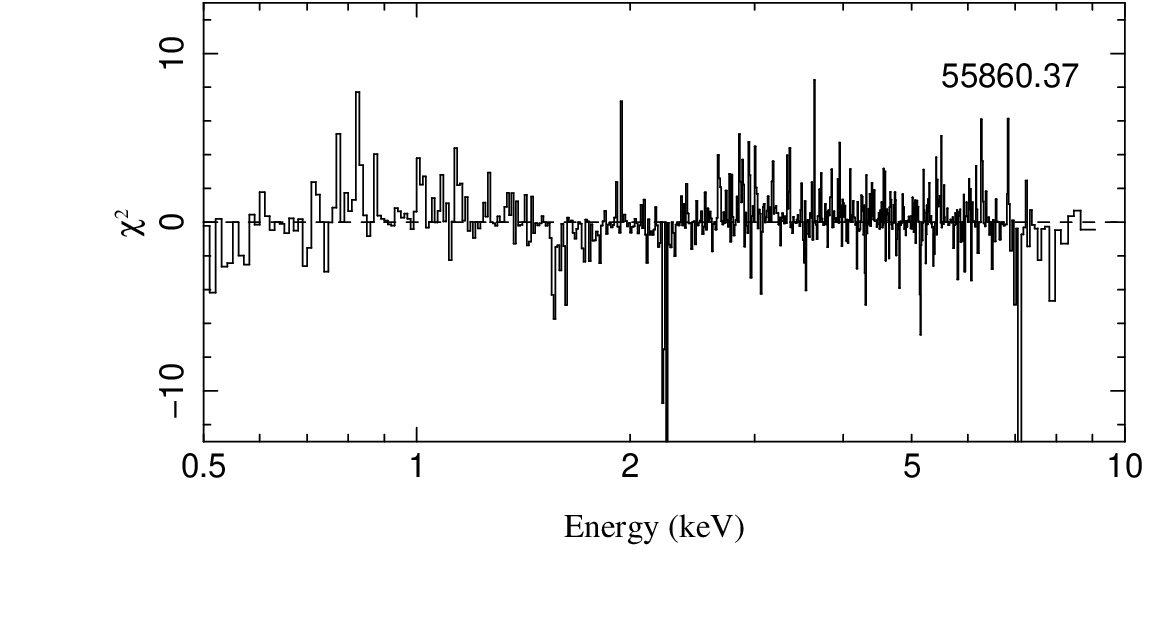}
 } &
 {\centering
 \includegraphics[width=52mm, height=21mm, trim = 32mm 08mm 3mm 0mm, clip]{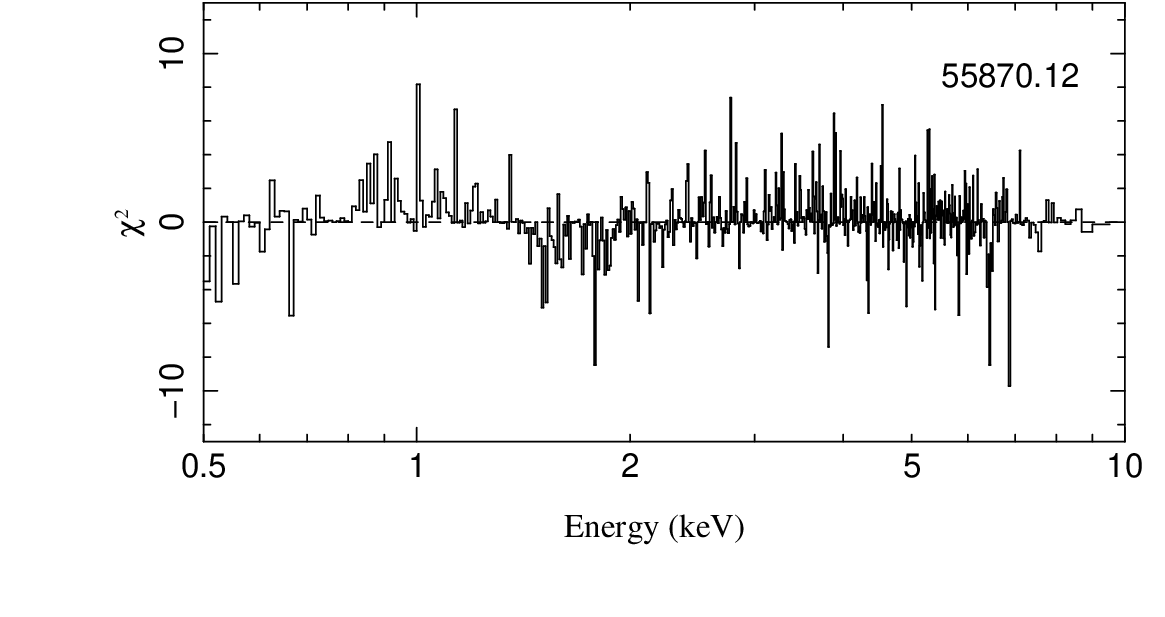}
 } \\
 \end{tabular}
 \caption{
 Residuals to the MCMC model with minimum \csq for the first 24
 observations. The MJD of the observation is indicated in the top-right
 of every plot.
 }
 \label{f:mcmc_res1}
\end{figure}

The MCMC simulations point to a spectral hardening factor ranging between
1.9--2.1 for the minimum \csq\ model. While this value is somewhat higher
than the generally accepted value of \about1.7, recent works, e.g. by
\citet{ReynoldsMiller2013} and \citet{Salvesen+2013}, have reported
sources where a higher \hd\ was required by the data. Additionally
we note from the \csq\ analysis that this `acceptable' range of \hd\
moves from low distances for low \mbh\ to high distances for high \mbh\
(see. e.g the heat maps in Figs.~\ref{f:heat-5m}--\ref{f:heat-15m}). When
an independent measurement of either the accretor's mass or the system
distance is available in the future, the above correlation can be used
to put a weak constraint on the other.

The fact that two extreme values of spin (maximal prograde for
higher inclination and maximal retrograde for lower inclinations) are
prefered in the traditional \csq\ treatment points to a degeneracy
in the fit-parameters.  Retrograde spin moves the ISCO outward,
and thus drops the temperature, but that is made up for by the high
color-correction factor.  Fewer photons (per unit area) are emitted by a
disk with an intrinsically lower $kT$, but the total photon count-rate
is compensated by the larger emitting area of a disk with larger inner
radius.  But in this case, the spectra clearly have high characterisic
$kT$ photons because increasing the emitting area alone is not sufficient
to obtain good fits; hence the need for a large color-correction factor.
This degeneracy is handled much better in MCMC simulations where even
though we searched the full range of possible spins, the near maximal
prograde spin is strongly favored over other scenarios, as seen in
Fig.~\ref{f:parcor} and Table~\ref{tab:mcmcglobalres}.

The MCMC solutions indicate a maximally spinning black hole, with a 90\%
confidence lower limit of 0.98. Even this lower limit is extremely
close to the canonical maximum value of \astar=0.998 calculated by
\citet{Thorne1974} for a geometrically thin, radiatively efficient
accretion disk. Other configurations however may allow higher
\astar. For example, black holes that harbor a thick, partially
pressure supported disk envisioned by \citet{Abramowicz+1978} might
have \astar$>$0.998. \citet{Sadowski+2011} have argued that the impact
of captured disk radiation by the black hole is neglgible at high
accretion rates, which can also push \astar\ beyond 0.998. See \S1.1 of
\citet{Sadowski+2011} for a summary of works by various authors on the
question of maximum spin of black holes. Since \src\ persistently accretes
in soft state, it is likely that it accretes at a significant fration of
the Eddington rate. Although absence of strong disk winds may disfavor an
\about Eddington or super-Eddington scenario.  Thus while extremely high
spins are not physically implausible, we note that deviations of the real
accretion disk from the theoretical model we used, and/or some calibration
uncertainty may also lead to high-spin solutions. Given that \src\ has
been observed using multiple X-ray missions and recent data from all of
these point to high spin \citep[see, e.g.,][]{Nowak+2008,Nowak+2012},
calibration related  errors are likely small.

The \swift\ X-ray spectra of \src\ are strongly dominated by thermal
photons indicating that the source is in a soft state.  Most other
X-ray observations, e.g. as presented in \citet{NowakWilms1999},
\citet{Wijnands+2002}, \citet{Nowak+2008}, \citet{Nowak+2012} also
suggest that the source is predominantly in a soft state. Furthermore,
low variability in the Fourier power spectra of the source
\citep{Wijnands+2002, Nowak+2008} and very low upper limits on any radio
emission \citep{Russell+2011} also point towards a persistent soft state
for this source. It is known that the luminosity of X-ray binaries in soft
state are typically $\gtrsim$few percent of their Eddington luminosity
\citep{Maccarone2003}. In Figs.~\ref{f:heat-5m}--\ref{f:heat-15m} the soft
state \lledd\ range is encompassed by yellow, green, and bluish colors.

While we expect \lledd\ for \src\ to be greater than a few percent,
it is unlikely that \lledd\about1. This is because luminosities close
to Eddington would drive strong winds that are not seen in this system.
Based on the heat maps of \lledd, smaller distances ($\sim$5 kpc) are
not favored because the best-fit luminosities are too low.  Also, higher
accretor masses would require higher distances for the luminosity to be in
the comfort zone for a soft state X-ray binary. Independent observations
of the strength of the ISM absorption lines in the direction of \src\
\citep{Nowak+2008, Yao+2008} also require the distance to be greater
than 5 kpc.

Our modeling of the \swift\ data does not constrain the masses of the
binary components, and we have assumed that the accretor is a black
hole based on its X-ray spectral and temporal properties.  On the other
hand, \citet{Bayless+2011} \citep[also see][]{Mason+2012} have recently
proposed a neutron star accretor for this system based on modeling the
optical light curve.  A thin, axisymmetric, uneclipsed disk produces a
constant flux in their models, and the optical modulation is assumed to
be entirely due to X-ray heating of the donor star.  Given the lack of
eclipses in the optical light curve, their models do not constrain the
orbital inclination or mass ratio very strongly.  The models weakly prefer
a mass ratio in the range of 0.025--0.3. Since neutron stars are less
massive than black holes, they therefore suggest a neutron star accretor.
However, the true structure of the disk may be more complicated.  In cases
of Accretion Disk Corona, optical orbital variability is associated with
the disk.  E.g., the disk-rim is raised where the incoming accretion
stream interacts with the disk. Partial optical eclipses may introduce
further orbital modulation. Additional observational evidence for the
disk's optical variability comes from studies of the long-term correlation
between the optical and X-ray light curves by \citep{Russell+2010} whose
results favor a disk origin (either via viscous or X-ray reprocessing)
for the optical light.

To summarize, the main conclisions of our joint analyses of the {\em
Swift} observations are as follows:\\
(1) The simulations suggest that the orbital inclination of the X-ray 
emitting inner accretion disk is \inclval\ degrees.\\
(2) The average column density towards the source, including extrinsic
(i.e. the ISM, and constant in time) and intrinsic (within the binary,
presumably from a disk-wind, and potentially time-variable) column to
be \nhval\ for the observations we have analyzed.\\
(3) The black hole spin is prograde and near maximal. Our MCMC simulations
indicate a 90\% confidence lower limit on the value of \astar\ to be 0.98.\\
(4) The system is located at a distance of $>$5 kpc, and possibly
farther than 10 kpc if \mbh$>$10\msun.

In addition to the new constraint on the inclination, the results presented 
here (based on data obtained using a modest observatory like {\em Swift}),
are consistent with, and strengthen the previous constraints that were made
not only using X-rays but also optical. 
While presenting new observations of this source, these 
results demonstrate the capabilities of long-term monitoring campaigns
to provide new insights and constrain accretion physics using the 
{\em Swift} mission.
% /*}}}*/

% Acknowledgements and facilities /*{{{*/
\acknowledgments
We thank the anonymous reviewers for suggestions that have greatly
improved the paper.  It is a pleasure to thank the \swift\ team for
coordinating the observations.  DM would also like to thank Mateusz
Ruszkowski and Jason Goodman for accomodating our many requests to use
the {\it galaxy} and {\it zephyr} clusters respectively. Thanks also
to Greg Salvesen for a discussion about black hole binaries with spins
misaligned w.r.t. the orbit.  This research made extensive use of data
obtained from the HEASARC data archive, provided by NASA's Goddard Space
Flight Center, and NASA's Astrophysics Data System. We thank NASA/\swift\
for funding this research.

{\it Facility:} \facility{\swift\ (XRT)}
% /*}}}*/

  % /*}}}*/ 

% Tables   /*{{{*/

% Table: Observation log /*{{{*/
\begin{deluxetable}{lccccccc}
\tabletypesize{\scriptsize}
\tablecaption{Observation log and fits using phenomenological 
{\it diskbb}+{\it power law} model (assuming 
\nh=1.5$\times$10$^{21}$ cm$^{-2}$).
\label{tab:obslog}
}
\tablewidth{0pt}
\tablehead{
\colhead{ObsID} & 
\colhead{\swift\ observation start} & 
\colhead{XRT exp.} & 
\colhead{kT$_{\rm dbb;in}$} & 
\colhead{N$_{\rm dbb}$} & 
\colhead{$\Gamma$} &
\colhead{N$_{\rm pl}$} &
\colhead{$\chi^2/\nu$} \\
 \colhead{} & 
 \colhead{MJD} & 
 \colhead{time (s)} & 
 \colhead{(keV)} & 
 \colhead{} & 
 \colhead{} &
 \colhead{} &
 \colhead{}
}
\startdata
00030959001 & 54282.549901 (2007-07-01 13:08:01) &   3917 & $1.45_{-0.03}^{+0.02}$ & $9.20_{-0.5}^{+0.5}$ & $1.9_{-0.1}^{+0.2}$ & $0.029_{-0.004}^{+0.004}$ & 698.0/597\\
00030959002 & 55516.216502 (2010-11-16 05:08:01) &   1699 & $1.47_{-0.05}^{+0.04}$ & $11.7_{-0.9}^{+1.2}$ & $1.9_{-0.4}^{+0.8}$ & $0.015_{-0.008}^{+0.008}$ & 528.6/480\\
00030959003 & 55519.025153 (2010-11-19 00:31:01) &   1610 & $1.44_{-0.05}^{+0.05}$ & $11.3_{-1.0}^{+1.3}$ & $1.6_{-0.2}^{+0.3}$ & $0.020_{-0.006}^{+0.006}$ & 583.1/524\\
00030959004 & 55522.903886 (2010-11-22 21:37:01) &   1649 & $1.39_{-0.03}^{+0.02}$ & $12.5_{-0.8}^{+0.9}$ & $2.2_{-0.5}^{+0.9}$ & $0.011_{-0.006}^{+0.006}$ & 551.8/499\\
00030959005 & 55525.578465 (2010-11-25 13:48:01) &   1624 & $1.44_{-0.04}^{+0.02}$ & $11.1_{-0.4}^{+1.0}$ & $0.5_{-0.5}^{-0.5}$ & $0.000_{-0.000}^{+0.003}$ & 564.0/498\\
00091070001 & 55700.488761 (2011-05-19 11:40:01) &   3118 & $1.55_{-0.02}^{+0.02}$ & $11.1_{-0.5}^{+0.5}$ & $2.1_{-0.2}^{+0.2}$ & $0.026_{-0.005}^{+0.006}$ & 746.8/627\\
00091070002 & 55710.053303 (2011-05-29 01:13:01) &   3199 & $1.39_{-0.03}^{+0.03}$ & $14.4_{-1.0}^{+1.2}$ & $1.5_{-0.1}^{+0.2}$ & $0.025_{-0.004}^{+0.004}$ & 666.2/611\\
00091070003 & 55720.095701 (2011-06-08 02:14:01) &   2759 & $1.51_{-0.04}^{+0.04}$ & $13.3_{-0.9}^{+1.2}$ & $1.4_{-0.3}^{+0.3}$ & $0.013_{-0.005}^{+0.005}$ & 625.7/618\\
00091070004 & 55730.049886 (2011-06-18 01:08:01) &   2654 & $1.60_{-0.03}^{+0.02}$ & $11.7_{-0.5}^{+0.6}$ & $1.8_{-0.4}^{+0.7}$ & $0.008_{-0.005}^{+0.005}$ & 688.7/626\\
00091070005 & 55740.427032 (2011-06-28 10:12:01) &   3278 & $1.59_{-0.03}^{+0.03}$ & $11.5_{-0.6}^{+0.7}$ & $1.7_{-0.1}^{+0.1}$ & $0.044_{-0.006}^{+0.006}$ & 812.4/662\\
00091070006 & 55750.513072 (2011-07-08 12:15:01) &   3029 & $1.59_{-0.03}^{+0.02}$ & $11.7_{-0.6}^{+0.6}$ & $2.0_{-0.2}^{+0.3}$ & $0.026_{-0.007}^{+0.007}$ & 672.7/611\\
00091070007 & 55760.682514 (2011-07-18 16:19:01) &   1819 & $1.54_{-0.02}^{+0.02}$ & $12.0_{-0.8}^{+0.9}$ & $2.4_{-0.3}^{+0.4}$ & $0.028_{-0.008}^{+0.009}$ & 690.3/561\\
00091070008 & 55763.040674 (2011-07-21 00:55:01) &   1624 & $1.47_{-0.04}^{+0.03}$ & $12.2_{-0.8}^{+1.0}$ & $1.8_{-0.2}^{+0.3}$ & $0.026_{-0.007}^{+0.007}$ & 609.0/547\\
00091070009 & 55770.593621 (2011-07-28 14:11:01) &   3149 & $1.44_{-0.03}^{+0.02}$ & $12.5_{-0.7}^{+0.8}$ & $1.7_{-0.2}^{+0.3}$ & $0.013_{-0.004}^{+0.004}$ & 714.1/599\\
00091070010 & 55780.356819 (2011-08-07 08:30:01) &   3149 & $1.57_{-0.03}^{+0.02}$ & $13.0_{-0.6}^{+0.7}$ & $1.7_{-0.2}^{+0.2}$ & $0.027_{-0.006}^{+0.006}$ & 785.4/650\\
00091070011 & 55791.394460 (2011-08-18 09:24:01) &   3224 & $1.64_{-0.03}^{+0.02}$ & $13.5_{-0.6}^{+0.6}$ & $1.9_{-0.3}^{+0.5}$ & $0.020_{-0.007}^{+0.008}$ & 759.6/651\\
00091070012 & 55800.357926 (2011-08-27 08:32:01) &   3100 & $1.63_{-0.02}^{+0.02}$ & $10.9_{-0.5}^{+0.5}$ & $2.2_{-0.2}^{+0.3}$ & $0.021_{-0.005}^{+0.006}$ & 741.0/646\\
00091070013 & 55810.184187 (2011-09-06 04:22:01) &   2458 & $1.46_{-0.03}^{+0.02}$ & $13.4_{-0.7}^{+0.7}$ & $2.0_{-0.2}^{+0.3}$ & $0.026_{-0.006}^{+0.006}$ & 657.5/584\\
00091070014 & 55820.417581 (2011-09-16 09:58:00) &   3069 & $1.47_{-0.02}^{+0.02}$ & $11.8_{-0.5}^{+0.6}$ & $2.1_{-0.1}^{+0.2}$ & $0.033_{-0.005}^{+0.005}$ & 757.0/603\\
00091070015 & 55830.191125 (2011-09-26 04:32:01) &   3279 & $1.45_{-0.02}^{+0.02}$ & $12.3_{-0.6}^{+0.6}$ & $1.9_{-0.2}^{+0.2}$ & $0.026_{-0.005}^{+0.005}$ & 867.6/606\\
00091070016 & 55840.227257 (2011-10-06 05:23:01) &   3168 & $1.54_{-0.02}^{+0.02}$ & $12.2_{-0.5}^{+0.6}$ & $2.3_{-0.2}^{+0.3}$ & $0.025_{-0.006}^{+0.005}$ & 843.7/632\\
00091070017 & 55850.390926 (2011-10-16 09:19:01) &   3189 & $1.42_{-0.03}^{+0.02}$ & $14.3_{-0.7}^{+0.8}$ & $1.8_{-0.1}^{+0.2}$ & $0.026_{-0.005}^{+0.005}$ & 895.8/611\\
00091070018 & 55860.365026 (2011-10-26 08:42:01) &   2834 & $1.52_{-0.02}^{+0.02}$ & $12.8_{-0.6}^{+0.6}$ & $2.3_{-0.2}^{+0.3}$ & $0.022_{-0.005}^{+0.006}$ & 802.7/616\\
00091070019 & 55870.124516 (2011-11-05 02:56:01) &   3209 & $1.56_{-0.03}^{+0.02}$ & $11.3_{-0.5}^{+0.6}$ & $1.8_{-0.2}^{+0.2}$ & $0.027_{-0.006}^{+0.005}$ & 810.9/638\\
00091070020 & 55880.144007 (2011-11-15 03:23:01) &   3069 & $1.51_{-0.02}^{+0.02}$ & $13.2_{-0.6}^{+0.6}$ & $2.0_{-0.2}^{+0.3}$ & $0.017_{-0.005}^{+0.005}$ & 757.9/613\\
00091070021 & 55890.318844 (2011-11-25 07:35:00) &   3172 & $1.39_{-0.02}^{+0.02}$ & $13.2_{-0.7}^{+0.7}$ & $2.0_{-0.2}^{+0.3}$ & $0.018_{-0.005}^{+0.005}$ & 645.6/554\\
\enddata
\end{deluxetable} % End of table/*}}}*/

% Table: Global MCMC results /*{{{*/
\begin{deluxetable}{ccc}
\tabletypesize{\scriptsize}
\tablecaption{Global spectral parameters, i.e. parameters that were not
allowed to change between observations, from MCMC simulations.
\label{tab:mcmcglobalres}
}
\tablewidth{0pt}
\tablehead{
\colhead{Parameter} & 
\colhead{Minimum \csq\ model} & 
\colhead{90\% confidence interval}\\
}
\startdata
$i$ (\degr) & 77.6 & (75.4, 79.1) \\
\nh\ (10$^{21}$ cm$^{-2}$) & 1.22 & (1.16, 1.25) \\
\astar & --- & $>0.98$ \\
\enddata
\end{deluxetable} % End of table/*}}}*/

% Table: Local MCMC results /*{{{*/
\begin{deluxetable}{lcccc}
\tabletypesize{\scriptsize}
\tablecaption{Local spectral parameters, i.e. parameters that changed from
one observation to another, from MCMC simulations. The values of \mdot$_{disk}$
are in units of 10$^{18}$ $g/s$.
\label{tab:mcmclocalres}
}
\tablewidth{0pt}
\tablehead{
\colhead{ObsID} & 
\colhead{\mdot$_{disk}$ for min. \csq\ model} & 
\colhead{90\% confidence interval for \mdot$_{disk}$} &
\colhead{\hd\ for min. \csq\ model} & 
\colhead{90\% confidence interval for \hd}
}
\startdata
00030959001 & 0.93 & (0.85, 3.48) &   1.98  & (1.62, 3.86)\\
00030959002 & 1.11 & (1.01, 4.22) &   2.00  & (1.66, 3.92)\\ 
00030959003 & 0.98 & (0.98, 4.04) &   2.01  & (1.68, 3.98)\\ 
00030959004 & 0.92 & (0.84, 3.46) &   1.96  & (1.61, 3.85)\\ 
00030959005 & 0.84 & (0.84, 3.41) &   2.07  & (1.71, 4.08)\\ 
00091070001 & 1.31 & (1.19, 4.94) &   1.96  & (1.61, 3.81)\\ 
00091070002 & 1.17 & (1.07, 4.41) &   1.92  & (1.64, 3.73)\\ 
00091070003 & 1.31 & (1.31, 5.31) &   1.98  & (1.69, 3.89)\\ 
00091070004 & 1.38 & (1.38, 5.58) &   2.02  & (1.71, 3.96)\\ 
00091070005 & 1.64 & (1.49, 6.21) &   1.95  & (1.59, 3.79)\\ 
00091070006 & 1.52 & (1.38, 5.74) &   1.96  & (1.62, 3.83)\\ 
00091070007 & 1.35 & (1.23, 5.11) &   1.89  & (1.56, 3.69)\\ 
00091070008 & 1.22 & (1.11, 4.59) &   1.95  & (1.61, 3.77)\\ 
00091070009 & 1.08 & (0.99, 4.10) &   1.96  & (1.63, 3.86)\\ 
00091070010 & 1.64 & (1.50, 6.24) &   1.93  & (1.58, 3.74)\\ 
00091070011 & 1.92 & (1.75, 7.28) &   1.92  & (1.64, 3.75)\\ 
00091070012 & 1.53 & (1.39, 5.78) &   2.00  & (1.70, 3.89)\\ 
00091070013 & 1.26 & (1.15, 4.75) &   1.88  & (1.55, 3.66)\\ 
00091070014 & 1.15 & (1.05, 4.27) &   1.89  & (1.56, 3.70)\\ 
00091070015 & 1.03 & (1.03, 4.19) &   1.92  & (1.63, 3.72)\\ 
00091070016 & 1.38 & (1.26, 5.21) &   1.92  & (1.58, 3.75)\\ 
00091070017 & 1.23 & (1.13, 4.67) &   1.86  & (1.54, 3.65)\\ 
00091070018 & 1.37 & (1.25, 5.13) &   1.90  & (1.63, 3.70)\\ 
00091070019 & 1.28 & (1.28, 5.18) &   1.97  & (1.61, 3.82)\\ 
00091070020 & 1.38 & (1.26, 5.22) &   1.92  & (1.58, 3.75)\\ 
00091070021 & 1.01 & (0.93, 3.76) &   1.89  & (1.57, 3.70)

\enddata
\end{deluxetable} % End of table/*}}}*/

% End -- tables  /*}}}*/

% Figures  /*{{{*/

 % /*}}}*/

 % /*}}}*/

 % /*}}}*/

 % /*}}}*/

 % /*}}}*/

 % /*}}}*/

 % /*}}}*/

 % /*}}}*/

 %/*}}}*/

\begin{figure} % MCMC bestfit residuals contd /*{{{*/
\centering
 \begin{tabular}{c c c }
 {\centering
 \includegraphics[width=52mm, height=21mm, trim = 10mm 08mm 3mm 0mm, clip]{resid_025.ps}
 } & 
 {\centering
 \includegraphics[width=52mm, height=21mm, trim = 32mm 08mm 3mm 0mm, clip]{resid_026.ps}
 } & 
 \\
 \end{tabular}
 \caption{
 Residuals to the MCMC model with minimum \csq for the last two
 observations. The MJD of the observation is indicated in the top-right
 of every plot.
 }
 \label{f:mcmc_res2}
\end{figure} %/*}}}*/

\begin{figure}
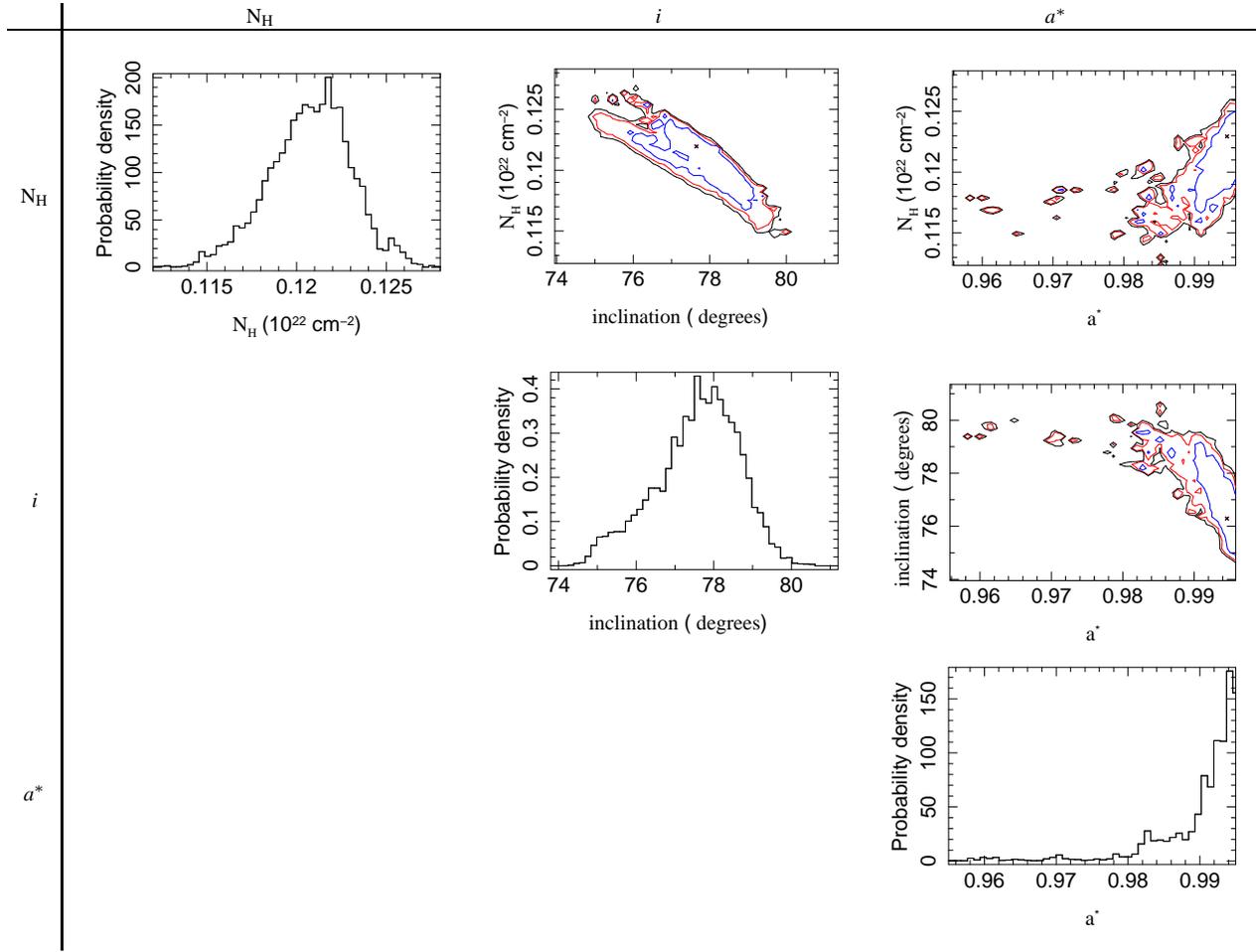
 % MCMC /*{{{*/
% nh = p0 ; a*=p1 ; i=p2 ; mbh=p3 ; d=p5 
\centering
 \begin{tabular}{c | c c c }
  & \nh & $i$ & \astar \\
 \hline
 & & & \\
 \nh     & \pxpy{0}{0}{4.9} & \pxpy{2}{0}{4.9} & \pxpy{1}{0}{4.9} \\
 & & & \\
 $i$     &                  & \pxpy{2}{2}{4.9} & \pxpy{1}{2}{4.9} \\
 & & & \\
 \astar  &                  &                  & \pxpy{1}{1}{4.9} \\
 & & & \\
 \end{tabular}
 \caption{
 MCMC results from a chain of 5,757,000 elements after rejecting data
 from the initial burn-in period.  The marginalized 1D histograms along
 the diagonal panels clearly show a single-peaked distribution for \nh\
 and $i$. On the other hand we can only obtain a lower limit on the
 spin parameter. The off-diagonal contour plots show the correlation
 between \nh, $i$, and \astar.  For the contour plots, the blue, red,
 and black colors correspond to 68\%, 90\%, and 95\% confidence contours.
 }
 \label{f:parcor}
\end{figure} %/*}}}*/

% End -- figures  /*}}}*/

\end{document}